\pdfoutput=1
\documentclass[smallextended]{svjour3}

\def \toolname {\textit{PTM4Tag+ }}
\def \toolnamenospace {\textit{PTM4Tag+}}

\usepackage[normalem]{ulem}
\usepackage[utf8]{inputenc}
\usepackage{amsmath}
\usepackage{graphicx}
\usepackage{comment}
\usepackage{caption}
\usepackage{multirow}
\usepackage{enumitem}
\usepackage{array}
\usepackage{tcolorbox}
\usepackage{xcolor}
\usepackage{balance}
\usepackage{mathtools}
\usepackage{url}
\usepackage{graphicx}
\usepackage{cite}
\usepackage{natbib}
\usepackage{microtype}

\begin{document}
\title{
\toolnamenospace: Tag Recommendation of Stack Overflow Posts with Pre-trained Models}

\author{Junda He, Bowen Xu, Zhou Yang, DongGyun Han, Chengran Yang, Jiakun Liu, Zhipeng Zhao, and David Lo}

\institute{J. He, Z. Yang, C. Yang, J. Liu, Z. Zhao, and D. Lo
\at School of Computing and Information Systems, Singapore Management University, Singapore\\ 
\email{\{jundahe,zyang,cryang,jkliu,zpzhao,davidlo\}@smu.edu.sg}
\and
B. Xu \at 
North Carolina State University, United States
 \email{bxu22@ncsu.edu}
           \and
        D. Han \at
Department of Computer Science, Royal Holloway, University of London Egham, UK \\
        \email{DongGyun.Han@rhul.ac.uk}
        }
\date{Received: date / Accepted: date}

\authorrunning{He. et al.}
\titlerunning{\toolnamenospace}

\maketitle
\begin{abstract}
Stack Overflow is one of the most influential Software Question \& Answer (SQA) websites, hosting millions of programming-related questions and answers. Tags play a critical role in efficiently organizing the contents in Stack Overflow and are vital to support a range of site operations, e.g., querying relevant content. Poorly selected tags often raise problems like tag ambiguity and tag explosion. Thus, a precise and accurate automated tag recommendation technique is demanded.

Inspired by the recent success of pre-trained models (PTMs) in natural language processing (NLP), we present \toolnamenospace, a {\em tag recommendation framework} for Stack Overflow posts that utilizes PTMs in language modeling. \toolname is implemented with a triplet architecture, which considers three key components of a post, i.e., Title, Description, and Code, with independent PTMs. 
We utilize a number of popular pre-trained models, including the BERT-based models (e.g., BERT, RoBERTa, CodeBERT, BERTOverflow, and ALBERT), and encoder-decoder models (e.g., PLBART, CoTexT, and CodeT5).  
Our results show that leveraging CodeT5 under the \toolname framework achieves the best performance among the eight considered PTMs and outperforms the state-of-the-art Convolutional Neural Network-based approach by a substantial margin in terms of average $Precision@k$, $Recall@k$, and $F1$-$score@k$ (k ranges from 1 to 5). Specifically, CodeT5 improves the performance of $F1$-$score@1$-$5$ by 8.8\%, 12.4\%, 15.3\%, 16.4\%, and 16.6\%. Moreover, to address the concern with inference latency, we experiment \toolname  with smaller PTM models (i.e., DistilBERT, DistilRoBERTa, CodeBERT-small, and CodeT5-small). We find that although smaller PTMs cannot outperform larger PTMs, they still maintain over 93.96\% of the  performance on average, meanwhile shortening the mean inference time by more than 47.2\%.

\end{abstract}
\section{Introduction}
\label{sec:intro}
Stack Overflow (SO) is the largest online Software Question \&
Answer (SQA) platform that facilitates collaboration and communications among developers in a wide range of programming-related activities. As of January 2023, Stack Overflow has more than 20 million registered users and hosted over 23 million questions with 34 million answers.\footnote{\url{ https://stackexchange.com/sites?view=list##traffic} } 
Stack Overflow has accumulated extensive resources for assisting software developers in their daily development process.

The rapid growth of Stack Overflow highlights the need to manage the site's contents at a large scale. To address this challenge, Stack Overflow uses \textit{tags} to categorize and structure the questions. Tags describe the topics and provide a concise summary of the question. Selecting accurate and appropriate tags can help many aspects of the site usage, e.g., connecting the expertise among different communities, triaging questions to the appropriate set of people with the right expertise, and assisting users in searching related questions~\citep{deeptagrec,tagcnn,tagcombine}. However, tags on the site are constructed through the process known as folksonomy,\footnote{\url{https://en.wikipedia.org/wiki/Folksonomy}} where the majority of tags are generated by users.
The quality of tags highly depends on users' level of expertise, English skills, etc. Tags selected by different users are likely to be inconsistent, triggering problems like tag ambiguity (i.e., the same tag is used for various topics)~\citep{deeptagrec} and tag explosion (i.e.,  multiple tags are used for the same topic)~\citep{barua2014developers}. As a result, users are hindered from finding relevant questions, and the productivity of programmers is decreased. These negative impacts motivate researchers to develop an automated tag recommendation technique to recommend high-quality tags for questions.



In this paper, we characterize the task of tagging SO posts as a multi-label classification problem following previous literature~\citep{entagrec, tagcnn, post2vec}, i.e., selecting the most relevant subset of tags from a large group of tags. Tagging SO posts is considered challenging for several reasons. First, the posts in Stack Overflow cover an extensive range of topics, resulting in an extremely large set of tags (i.e., over 10 thousand available tags). Second, allowing users to freely tag their posts introduces great inconsistencies and makes the tag set rather sparse.
It is non-trivial to build one model to accurately capture the semantics of the posts and establish connections between the posts and the related tags.

A growing body of literature tackled the \textit{tag recommendation task} of SO posts ~\citep{tagcnn,post2vec}. The state-of-the-art solution for the task is Post2Vec~\citep{post2vec}. Post2Vec is a deep learning-based tag recommendation approach using Convolutional Neural Networks (CNNs)~\citep{schmidhuber2015deep} as the feature extractors. Motivated by the success of CNN in the SO post tag recommendation task, we focus on further improving the tag recommendation performance by leveraging the transformer-based pre-trained models (PTMs). Compared to Post2Vec,
PTMs enhance CNN with the self-attention mechanism and provide a much better model initialization with pre-training knowledge. Such properties make PTMs suitable for tagging SO posts and prompt our interest in adopting them.

Transformer-based PTMs have achieved phenomenal performance in the Natural Language Processing (NLP) domain~\citep{bert}\citep{roberta}\citep{t5}. They are shown to vastly outperform other techniques like LSTM~\citep{bilstem} and CNN~\citep{schmidhuber2015deep}. Inspired by their success, there is an increased interest in applying PTM to the field of software engineering (SE).
Such PTMs are proven to be effective in multi-class or pair-wise text classification tasks such as sentiment analysis~\citep{zhang2020sentiment} and API review~\citep{chengran2022saner}. To the best of our knowledge, aside from our conference paper that this paper is extended from, no studies in the SE literature have investigated the performance of directly fine-tuning PTMs in handling a multi-label classification problem with thousands of labels. This motivates us to explore the effectiveness of PTMs in the task of SO tag recommendation.

Nonetheless, directly applying PTMs from the NLP-domain to SE-related downstream tasks has limitations. Texts in different domains (i.e., NLP and SE) usually have different word distributions and vocabularies. Software developers are free to create identifiers they prefer when writing code. The formulated identifiers are often compound words and arbitrarily complex, e.g., \textit{addItemsToList}~\cite{shi2022identifier}. 
Words may also have a different meaning for software engineering. For example, the word ``Cookie'' usually refers to a small chunk of data stored by the web browser to support easy access to websites for software engineers, and it does not refer to the ``baked biscuit'' in the context of SE. PTMs trained with natural language text may fail to capture the semantics of SE terminology. NLP-domain language models (e.g., BERT~\citep{bert} and RoBERTa~\citep{roberta}) are usually extensively pre-trained with a significantly larger corpus than the SE-domain PTMs as it is much more difficult to obtain a high-quality, large-scale corpus of a specialized domain. Since each kind of model has its potential strengths and weaknesses, it prompts our interest in exploring the impact and limitations of different PTMs from the NLP-domain and SE-domain.

In this paper, we introduce \toolnamenospace, a framework that utilizes popular pre-trained models and trains a multi-label classifier on the transformer architecture for recommending tags for SO posts. We evaluate the performance of \toolname with different PTMs to identify the best variant. We categorize PTMs from two dimensions: domain-difference (NLP-domain vs. SE-domain) and architecture-difference (encoder-only vs. encoder-decoder). While the encoder-only BERT-based PTMs have been widely used in language modeling, we delve into the language representation from another family of pre-trained models: the encoder-decoder models. Encoder-decoder models like CodeT5 \citep{codet5} and PLBART \citep{plbart} boost the state-of-art-performance for numerous SE-related downstream tasks (e.g., code summarization, code generation, and code translation). 
Intuitively, encoder-decoder models should also be powerful in language modeling. Recent studies have demonstrated that the embeddings generated by T5 are superior to those generated by BERT~\citep{sentencet5}. Nonetheless, in the SE domain, there are limited studies on leveraging popular encoder-decoder models (e.g., CodeT5 and PLBART) in classification tasks and evaluating the representation generated by these models. To fill this gap, we examine the efficacy of popular encoder-decoder models under the \toolname framework.

Furthermore, in order to increase the usability of \toolnamenospace, we reduce the size of \toolname by adopting smaller PTMs. In terms of deployment, a tool with a smaller size is more practical for integration in modern applications as it requires less storage and runtime memory consumption. On the other hand, the inference latency would be much faster, and user satisfaction can be hugely increased in the context of tagging SO posts. To address the concern with model size, we first identify several PTMs that can yield promising results under the \toolname framework and then leverage the smaller version of these PTMs to train four more variants of \toolname.

To obtain a comprehensive understanding of \toolnamenospace, we answer the following research questions:

\textbf{\textit{RQ1: Out of the eight variants of \toolname with different PTMs, which gives the best performance?}}
Considering that each model has its potential strengths and weaknesses, it motivates us to study the impact of adopting different PTMs in \toolnamenospace. Namely, we compare the results of BERT~\citep{bert}, RoBERTa~\citep{roberta}, ALBERT~\citep{albert}, CodeBERT\citep{CodeBERT}, BERTOverflow~\citep{bertoverflow}), PLBART~\citep{plbart}, CoTexT~\citep{cotext}, and CodeT5~\citep{codet5}).

\textbf{\textit{RQ2: How is the performance of \toolname compared to the state-of-the-art approach in Stack Overflow tag recommendation?}}
In this research question, we examine the effectiveness and performance of \toolname by comparing it with the CNN-based state-of-the-art baseline for the tag recommendation task of Stack Overflow, i.e., Post2Vec~\citep{post2vec}.

\textbf{\textit{RQ3: Which component of post benefits \toolname the most?}}
\toolname is implemented with a triplet architecture, which encodes the three components of an SO post, i.e., Title, Description, and Code with different Transformer-based PTMs. Considering that each component may carry a different level of importance for the tag recommendation task, we explore the contribution of each component by conducting an ablation study.

\textbf{\textit{RQ4: How is the performance of \toolname with smaller pre-trained models?}}
Utilizing PTMs gives great performance but results in high inference latency and difficulties in deployment~\citep{compressor}. In this RQ, we select smaller PTMs to train \toolnamenospace. 

As an extended version of our previous work \citep{ptm4tag}, which proposed and evaluated the \textit{PTM4Tag} framework, this paper further enhances \textit{PTM4Tag} by empirically experimenting with more PTMs, including a set of Seq2Seq models and smaller PTMs.

We derived several useful findings from the experiments and we highlight the difference to our previous work in \textbf{bold}: 

\begin{enumerate}[leftmargin=*]

\item \textbf{\toolname with CodeT5 outperforms other PTMs and the previous state-of-the-art approach (i.e. Post2Vec) by a large margin.} 

\item Although \toolname with ALBERT and BERTOverflow perform worse than Post2Vec, other PTMs are capable of outperforming Post2Vec. Thus, we conclude that leveraging PTMs can help to achieve promising results, but the PTM within \toolname needs to be rationally selected.

\item \textbf{Encoder-decoder PTMs can also benefit the \toolname framework. While encoder-decoder PTMs are usually used for generation tasks, we advocate future studies to include the Seq2Seq PTMs as baseline methods for representing SO posts.} 

\item \textbf{We reduce the size of \toolnamenospace. Specifically, the inference time is improved by more than 47.2\%, while the model preserves more than 93.96\% of the original performance.}

\end{enumerate}

The contributions of the paper are summarized as follows: 

\begin{enumerate}[leftmargin=*]

\item We propose \toolnamenospace, a transformer-based muli-label classifier, to recommend tags for SO posts. To the best of our knowledge, our work is the first to leverage pre-trained language models for tag recommendation of SO posts. Our proposed tool outperforms the previous state-of-the-art technique, Post2Vec, by a substantial margin. 

\item We explore the effectiveness of different PTMs by training eight variants of \toolname and comparing their performance. We consider a set of BERT-based PTMs and \textbf{encoder-decoder PTMs}.

\item We further conducted an ablation study to investigate the contribution of each component to the task.

\item \textbf{We address the concern with model size and inference latency by involving a set of smaller PTMs under the \toolname framework. }
\end{enumerate}

The paper is structured as follows: Section \ref{sec:background} introduces the background knowledge for the tag recommendation task of SO, the state-of-the-art baseline approach, and five popular PTMs that are investigated in our study. Section \ref{sec:methodology} describes our proposed approach in detail. Section \ref{sec:experiment} specifies the experimental settings. Section \ref{sec:results} presents the experimental results with analysis. In Section  \ref{sec:discussion}, we conducted a qualitative analysis, discussed the threats to validity and summarized our learned lessons. Section \ref{sec:related_work} reviews the literature on PTMs applied in SE, and the tag recommendation approaches for Software Question \& Answer sites. Finally, we conclude our work and discuss future work in Section \ref{sec:conclusion}.
\section{Background}
\label{sec:background}
In this section, we formalize the tag recommendation task for SO posts as a \textit{multi-label classification problem}. Then, we describe Post2Vec~\citep{post2vec}, the state-of-the-art tag recommendation approach that is used as a baseline in this paper. In the end, we introduce the pre-trained language models that are leveraged in \toolnamenospace. 

\subsection{Tag Recommendation Problem}
\label{sec:formal}
Considering that an SO post can be labeled by one or multiple tags, we regard the tag recommendation task for SO posts as a \textit{multi-label classification problem}.
We denote the corpus of SO posts by $\mathcal{X}$ and the collection of tags as $\mathcal{Y}$. Formally speaking, given an SO post $x \in \mathcal{X}$, the tag recommendation task aims to acquire a function $f$ that maps $x$ to a subset of tags $y = \{y_1, y_2, ..., y_l\}  \subset \mathcal{Y}$ that are most relevant to the post $x$. We denote the total number of training examples as $N$, the total number of available tags\footnote{\url{https://stackoverflow.com/help/tagging}} as $L$ and the number of tags of training data as $l$, such that $L = |\mathcal{Y}|$ and $l = | y |$. Note that one SO post can be labeled with at most five tags, so $l$ must be equal to or less than 5.\footnote{\url{https://resources.stackoverflow.co/topic/product-guides/topic-tag-targeting/}}

\subsection{Post2Vec}\label{sec:post2vec}
Xu et al. proposed Post2Vec~\citep{post2vec}, a deep learning-based tag recommendation approach for SO posts, which achieves state-of-the-art performance. Xu et al. trained several variants of Post2Vec to examine the architecture design from multiple aspects. In this paper, we select their best-performing variant as our baseline model. Specifically, our baseline model leverages CNN as the feature extractor of the post and divides the content of a post into three components, i.e., Title, Description, and Code. Each component of a post has its own component-specific vocabulary and is modeled separately with a different neural network.

\subsection{Pre-trained Language Models}

Recent trends in the NLP domain have led to the rapid development of transfer learning. Especially, substantial work has shown that pre-trained language models learn practical and generic language representations which could achieve outstanding performance in various downstream tasks simply by fine-tuning them on a smaller dataset, i.e., without training a new model from scratch~\citep{tracebert,jin2020bert,qu2019bert}. With proper training manner, the model can effectively capture the semantics of individual words based on their surrounding context and reflect the meaning of the whole sentence. 

\begin{figure}
	\centering
    \includegraphics[width=0.95\columnwidth]{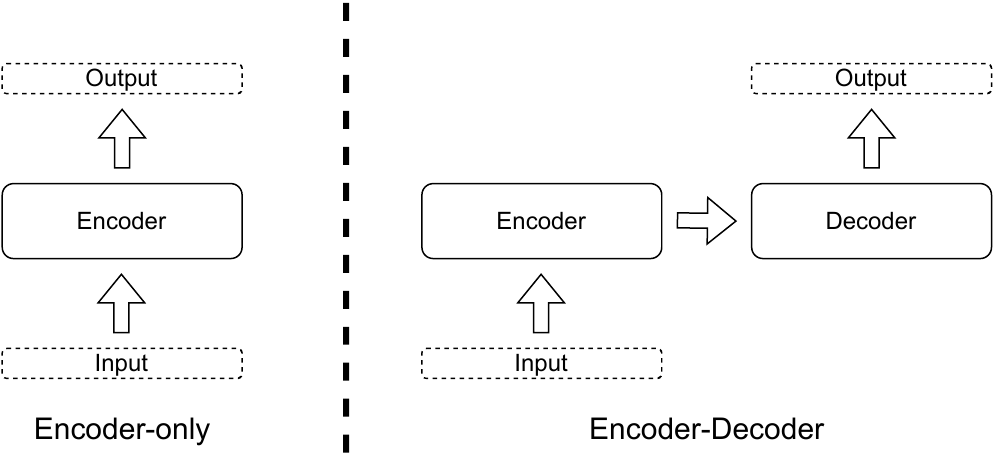}
	\caption{The architecture of encoder-only and encoder-decoder models}
	\label{fig:encoder}
\end{figure}

A major drawback of Post2Vec is that its underlying neural network (i.e., CNN) has limitations in modeling long input sequences. CNN requires large receptive fields to model long-range dependencies~\citep{schmidhuber2015deep}. However, increasing the receptive field dramatically reduces computational efficiency. We addressed this limitation by leveraging the Transformer-based PTMs, which enhanced the architecture with a self-attention mechanism and pre-trained knowledge obtained from other datasets. We categorized the considered PTMs in this paper into two types: the encoder-only PTMs and the encoder-decoder PTMs. The architectural difference between these two types of PTMs is illustrated in Fig \ref{fig:encoder}.
In Table \ref{tab:models}, we summarize the architecture, pre-training tasks, downstream tasks from the original paper, and language type of the PTMs used in this paper. Table \ref{tab:pretrain} and \ref{tab:abb} presents the details of the abbreviation used in Table \ref{tab:models}.

\subsubsection{BERT-based Pre-trained Models}
BERT-based Pre-trained Models utilize only the encoder stacks of a Transformer model. The attention layers have the ability to access every word of the input sentence, and these models are commonly referred to as having ``bi-directional" attention. The pre-training objectives of these models typically involve some form of corruption to a given sentence (i.e., mask language model and replaced token detection). BERT-based Pre-trained Models are powerful techniques for generating sentence embeddings and are well-suited for language understanding tasks. 

\begin{table}[]
\centering
\caption{Summarization of Pre-trained Models used in this Paper, including the pre-training task, architecture, downstream tasks, and language type}
\label{tab:models}
\resizebox{\textwidth}{!}{%
\begin{tabular}{c|c|c|c|c}
\hline
\textbf{Model} & \textbf{Training Tasks} & \textbf{Architecture} & \textbf{Downstream Tasks (SE tasks)} & \textbf{Language Type} \\ \hline
RoBERTa        & MLM                     & EN                    & -                                & NL                     \\ \hline
BERT           & MLM, NSP                & EN                    & -                                & NL                     \\ \hline
ALBERT         & MLM, NSP                & EN                    & -                                & NL                     \\ \hline
BERTOverflow   & MLM                     & EN                    & SER                              & NL                     \\ \hline
CodeBERT       & MLM, RTD                & EN                    & CR, CS QA, CT, BF, CSM           & NL, PL                 \\ \hline
PLBART         & DAE                     & ED                    & DD, CD, QA, CT, BF, CS, CG       & NL, PL                 \\ \hline
CodeT5         & SMLM, IT, IMLM, BDG     & ED                    & DD, CD, QA, CT, BF, CSM, CG      & NL, PL                 \\ \hline
CoTexT         & SMLM                    & ED                    & DD, BF CSM, CG                   & NL, PL                 \\ \hline
\end{tabular}%
}
\end{table}

\begin{table}[]
\centering
\caption{Description and Abbreviation of the Pre-training tasks mentioned in Table \ref{tab:models} }
\label{tab:pretrain}
\resizebox{\textwidth}{!}{%
\begin{tabular}{l|l|l}
\hline
Pre-train Tasks                                                     & Abb. & Description                                                                                                                                                                           \\ \hline
\begin{tabular}[c]{@{}l@{}}Mask Language \\ Modeling\end{tabular}   & MLM  & \begin{tabular}[c]{@{}l@{}}Given an input where certain tokens are hidden (masked), the model \\ predicts those missing tokens.\end{tabular}                                          \\ \hline
\begin{tabular}[c]{@{}l@{}}Next Sentence \\ Prediction\end{tabular} & NSP  & Assesses whether two provided sentences naturally appear in sequence.                                                                                                                 \\ \hline
\begin{tabular}[c]{@{}l@{}}Replaced Token \\ Detection\end{tabular} & RTD  & \begin{tabular}[c]{@{}l@{}}Recognizes whether a specific token in the input is artificially generated, \\ rather than being from the original text.\end{tabular}                      \\ \hline
\begin{tabular}[c]{@{}l@{}}Denoising \\ Auto-Encoding\end{tabular}  & DAE  & \begin{tabular}[c]{@{}l@{}}From an input where tokens have been modified (through masking, deletion, \\ or replacement), the model aims to reproduce the original input.\end{tabular} \\ \hline
Seq2Seq MLM                                                         & SMLM & \begin{tabular}[c]{@{}l@{}}Using an encoder-decoder architecture, the model tries to predict a sequence \\ of tokens that have been masked out.\end{tabular}                          \\ \hline
Identifier Tagging                                                  & IT   & Classifies each token in the input based on whether it's an identifier or not.                                                                                                        \\ \hline
Identifier MLM:                                                     & IMLM & In the context of code, this task involves predicting masked-out identifiers.                                                                                                         \\ \hline
\begin{tabular}[c]{@{}l@{}}Bimodal Dual \\ Generation\end{tabular}  & BDG  & \begin{tabular}[c]{@{}l@{}}Given a natural language description/code, it produces \\ code/natural language description and vice versa.\end{tabular}                                   \\ \hline
\end{tabular}%
}
\end{table}

\begin{table}[]
\centering
\caption{Abbreviations of Architecture Type, Langauge Type, and Downstream Tasks in Table \ref{tab:models}}
\label{tab:abb}
\resizebox{\textwidth}{!}{%
\begin{tabular}{c|cc}
\hline
\multicolumn{1}{c|}{\textbf{Architecture Type}} & \multicolumn{2}{c}{\textbf{Downstream Tasks}}                                             \\ \hline
Encoder-only (EN)              & \multicolumn{1}{c|}{Software Entity Recognition (SER)} & Bug Fixing (BF)          \\ \hline
Encoder-decoder (ED)           & \multicolumn{1}{c|}{Code-to-Code Retrieval (CR)}       & Code Summarization (CSM) \\ \hline
\textbf{Language Type}                           & \multicolumn{1}{c|}{Code Search (CS)}                  & Defect Detection (DD)    \\ \hline
Natural Language (NL)          & \multicolumn{1}{c|}{Code Question Answering (QA)}      & Clone Detection (CD)     \\ \hline
Programming Language (PL)      & \multicolumn{1}{c|}{Code Translation (CT)}             & Code Generation (CG)     \\ \hline
\end{tabular}%
}
\end{table}

\subsubsection*{\textbf{BERT}}
BERT~\citep{bert} is based on the Transformer architecture~\citep{transformer} and contains the bidirectional attention mechanism. BERT is pre-trained on a large corpus of general text data, including the entire English Wikipedia dataset and the BooksCorpus~\citep{Zhu_2015_ICCV}. It has two pre-training tasks: Masked Language Modeling (MLM) and Next Sentence Prediction (NSP). Given an input sentence where some tokens are masked out, the MLM task predicts the original tokens for the masked positions. Given a pair of sentences, the NSP task aims to predict whether the second sentence in the pair is the subsequent sentence to the first sentence.

\subsubsection*{\textbf{DistilBERT}}
DistilBERT is a smaller and lighter version of the BERT model. As the number of parameters of PTMs is getting bigger and bigger, DistilBERT aims to reduce the computational cost of training and inference processes while preserving most of the performance of the larger models. In comparison with the BERT model, which has 12 hidden layers and 110 million parameters in total, DistilBERT only has six hidden layers and 66 million parameters. DistilBERT has leveraged knowledge distillation \citep{knowledge_distilation} during the pre-training phase and shown that it is possible to retain 97\% of BERT's language understanding abilities with a 40\% reduction in the model size and 60\% faster.

\subsubsection*{\textbf{RoBERTa}} RoBERTa~\citep{roberta} is mainly based on the original architecture of BERT, but modifies a few key hyper-parameters. It removes the NSP task and feeds multiple consecutive sentences into the model. RoBERTa is trained with a larger batch size and learning rate on a dataset that is an order of magnitude larger than the training data of BERT~\citep{bert,Zhu_2015_ICCV}.

\subsubsection*{\textbf{DistilRoBERTa}} 
Similar to DistilBERT, DistilRoBERTa is the distilled version of the RoBERTa model. By leveraging the same pre-training strategy (i.e., knowledge distillation), DistilRoBERTa has 34\% fewer parameters (i.e., 82 million parameters) than RoBERTa, but twice as fast.

\subsubsection*{\textbf{ALBERT}} ALBERT~\citep{albert} is claimed as \textbf{A} \textbf{L}ite \textbf{BERT}. 
ALBERT involves two parameter reduction techniques: factorized embedding parameterization and cross-layer parameter sharing. Factorized embedding parameterization breaks down the large embedding matrix into two small matrices, and cross-layer parameter sharing prevents the parameter from growing with the depth of the network. Both techniques can effectively reduce the number of parameters without compromising the performance. 
Additionally, it replaces the NSP task used by BERT with the Sentence Order Prediction (SOP) task. By doing so, ALBERT can significantly reduce the number of model parameters and facilitate the training process without sacrificing the model performance.

\subsubsection*{\textbf{CodeBERT}} CodeBERT follows the same architectural design as RoBERTa. However, CodeBERT \citep{CodeBERT} is pre-trained on both natural language (NL) and programming language (PL) data from the CodeSearchNet database~\citep{codesearchnet}. CodeBERT considers two objectives at the pre-training stage: Masked Language Modeling (MLM) and Replaced Token Detection (RTD). The goal of the RTD task is to identify which tokens are replaced from the given input. CodeBERT uses bimodal data (NL-PL pairs) as input at the pre-training stage to understand both forms of data. The CodeBERT model has been proven practical in various SE-related downstream tasks, such as Natural Language Code Search~\citep{zhou2021assessing,Huang2021CoSQA2W}, program repair~\citep{mashhadi2021apply}, etc~\citep{CodeBERT}.

Concurrent with the work by Feng et al., researchers from Huggingface have also released their own CodeBERT model,\footnote{\url{https://huggingface.co/huggingface/CodeBERTa-small-v1}} which is also pre-trained with the CodeSearchNet data. The CodeBERT model released by Feng et al. has the same number of layers and parameters as RoBERTa (12 layers and 128 million parameters). In comparison, HuggingFace's model only has six layers and 84 million parameters. In our paper, we refer to the model trained by Feng et al. as the \textbf{CodeBERT} model, and the model trained by HuggingFace as the \textbf{CodeBERT-small} model.


\subsubsection*{\textbf{BERTOverflow}} BERTOverflow ~\citep{bertoverflow}  is a SE-domain PTM trained with 152 million sentences from Stack Overflow. The author of BERTOverflow introduces a software-related named entity recognizer (SoftNER) that combines an attention mechanism with code snippets. The model follows the same design as the BERT model with 110 million parameters. Experimental results showed that leveraging embedding generated by BERTOverflow in SoftNER substantially outperformed BERT in the code and named entity recognition task for the software engineering domain. 


\subsubsection{Encoder-decoder Pre-trained Models}

In contrast with the BERT-based PTMs, which have an encoder-only architecture, encoder-decoder models follow the full transformer architecture.

\subsubsection*{\textbf{CodeT5}} Wang et al. proposed CodeT5 \citep{codet5}, which inherits the T5 (Text-To-Text Transfer Transformer) architecture and conducts the denoising sequence-to-sequence pre-training. To assist the model in better understanding programming languages, Wang et al. extend the denoising objective of T5 proposed with identifier tagging and prediction tasks. CodeT5 is trained to recognize which tokens are identifiers and to predict them from masked values. Moreover, CodeT5 leverages a  bimodal dual-generation task for NL-PL alignment. CodeT5 provides the ability for multi-task learning, and it is capable of being fine-tuned in numerous downstream tasks, including code summarization, code generation, code translation, code refinement, defect prediction, clone detection, etc.

Wang et al. implemented several versions of CodeT5 with different sizes, where CodeT5-small has 60 million parameters and
CodeT5-base has 220 million parameters. In this paper, we refer to CodeT5-base as \textbf{CodeT5} and CodeT5-small as \textbf{CodeT5-small}. 

\subsubsection*{\textbf{PLBART}} 
PLBART \citep{plbart} is proposed by Ahmad et al. and is capable of performing a broad spectrum of program understanding and generation tasks. PLBART inherits the BART \citep{bart} architecture and is trained using multilingual denoising tasks in Java, Python, and English.
PLBART is pre-trained with a large-scale bi-model corpus of both natural languages and programming languages, including Java, Python, and English. PLBART has been shown to outperform the rival state-of-the-art methods in code summarization, code translation, and code generation tasks. It is also capable of generating promising performance in a wide range of language understanding tasks, e.g., program repair, clone detection, and vulnerable code detection. 

\subsubsection*{\textbf{CoTexT}} Phan et al. introduced CoTexT(\textbf{Co}de and \textbf{Tex}t Transfer \textbf{T}ransformer) \citep{cotext}, which is also implemented in the T5 architecture. In the pre-training stage, CoText mainly leverages the bi-model data collected from the CodeSearchNet corpus \citep{codesearchnet} and GitHub repositories.  The effectiveness of CoTexT is demonstrated by evaluating four tasks, which are code summarization,  code generation, defect detection, and code refinement. Results show that CoText is capable of achieving better performance than CodeBERT and PLBART. 

\section{Methodology}
\label{sec:methodology} 
This section introduces our proposed tag recommendation framework for SO posts, \toolname in detail. The overall architecture of \toolname with three stages is illustrated in Figure~\ref{fig:architecture}.
The three stages are: \textbf{Pre-processing, Feature Extraction, and Classification}.
We first decompose each SO post into three components, i.e., \textit{Title}, \textit{Description}, and \textit{Code}.
Thus, \toolname is implemented with a \emph{triplet} architecture and leverages three PTMs as encoders to generate representations for each component.
Then, at the feature extraction stage, we feed the processed data into the used PTMs and represent each component as a feature vector.
The obtained feature vectors are then concatenated to construct the final representation of the SO post.
Finally, at the classification stage, the classifier maps the post representation to a tag vector that indicates the probability of each tag.

\subsection{Pre-processing}
During the pre-processing stage, we split an SO post into three components and then conduct tokenization. 
\begin{figure}
	\includegraphics[width=1\linewidth]{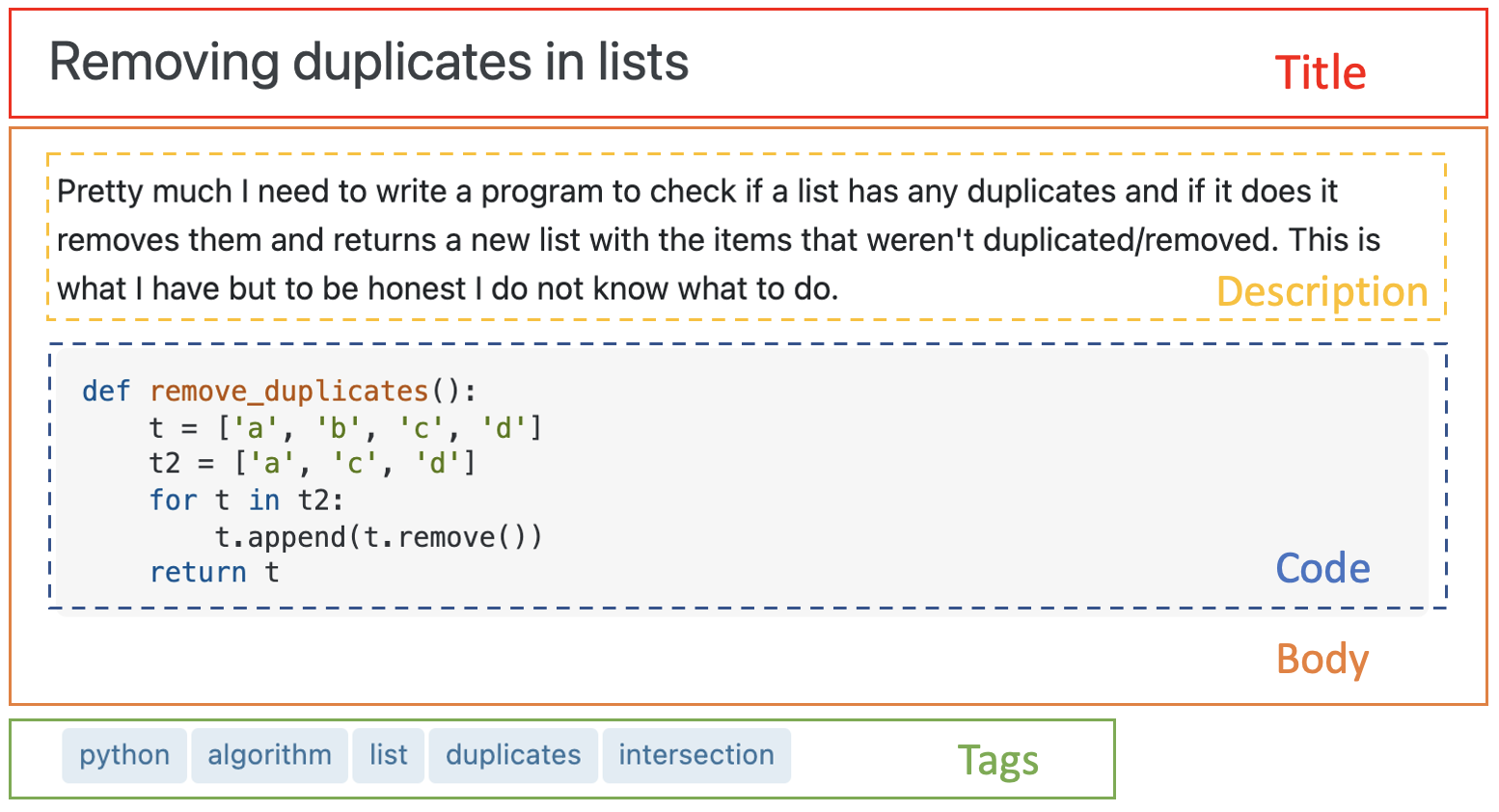}
	\vspace{-2mm}
	\caption{An example of an SO Post. A post contains a short title that summarizes the main content of this post. The body of a post can include detailed descriptions written in natural languages and code snippets.}
	\label{fig:post}
	\vspace{-4mm}
\end{figure}

\subsubsection{Post Component Extraction}
Figure~\ref{fig:post} illustrates a typical SO post that consists of three components: \textit{Title}, \textit{Body}, and \textit{Tags}.
The \textit{Title} summarizes the question, and the \textit{Body} provides details of the question that helps readers to understand the question.
Following prior studies \citep{SOTorrent}, we further divide the \textit{Body} into \textit{Description} blocks and \textit{Code} blocks. \textit{Description} blocks in a \textit{Body} are narrative sections that describe the context or problem in natural languages.
\textit{Code} refers to the parts of the post that are enclosed in HTML tags
\texttt{<pre><code>}. For simplicity, we refer to the sections enclosed in the \texttt{<pre><code>} HTML tags as \textit{code snippets} in this paper. However, the sections enclosed in the \texttt{<pre><code>} HTML tags may not always be actual code snippets; they can also be other types of text, like stack traces or error messages. 


Different from prior studies that discarded \textit{Code} blocks in Post during the pre-processing stage (because the \textit{Code} can be written by novices and have low quality), we keep \textit{Code} in our work.
This is because we observe that the \textit{Code} blocks in an SO post can provide valuable semantic information in recommending tags.
Take the SO post shown in Figure~\ref{fig:post} as an example. One of its tags is `{\tt python}', but neither \textit{Title} nor \textit{Description} explicitly mentions the post is {\tt python}-related.
If we only look at the title and description sections, it is unclear which programming language this post is asking about.
However, by adding the \textit{Code} into consideration, the grammar of Python can be easily used to infer that the post is likely to relate to the tag `{\tt python}'.
As a result, we consider that a post is made up of four components: the \textit{Title}, \textit{Description}, \textit{Code}, and \textit{Tags}. Our proposed \toolname framework takes \textit{Title}, \textit{Description}, and \textit{Code} as input and aims to predict the sets of tags that are most relevant to this post (i.e., the \textit{Tags}).

To decompose the SO posts into the above-mentioned four constituents, we identify the \textit{Title} and the \textit{Tag} of posts from the \texttt{Posts.xml} in the official data dump released by the Stack Overflow website.
To extract the \textit{Code} in posts, we use a regular expression 
$\texttt{<pre><code>}([\backslash s\backslash S]*?)\texttt{</code></pre>}$
to identify the code blocks in posts.
We then further remove the redundant HTML tags within these sections since these HTML tags are used for formatting and are irrelevant to the content of a post.

\subsubsection{Tokenization}
Since the design of \toolname leverages transformer-based PTMs, we rely on the corresponding tokenizer of the underlying pre-trained model to
generate token sequences.
The conventional PTMs usually accept a maximum input sequence length of 512 sub-tokens, which also always include two special tokens $\langle CLS\rangle$ and $\langle SEP\rangle$.
The $\langle CLS\rangle$ token (\textbf{CL}a\textbf{S}sification) is the first token of every input sequence, and the corresponding hidden state is then used as the aggregate sequence representation.
$\langle SEP\rangle$ token (\textbf{SEP}arator) is inserted at the end of every input sequence.
The problem of capturing long text arises since a significant proportion of the training samples has exceeded the maximum acceptable input limit of the PTMs.

By default, we tackle the problem using a \textit{head-only} truncation strategy~\citep{howtofineture}, which only considers the first 510 tokens (excluding the $\langle CLS\rangle$ and $\langle SEP\rangle$ tokens) as the input tokens. We also further discuss the
impact of the \textit{tail-only} truncation strategy in Section \ref{sec:results}. 
Instead of filtering less significant words, we apply the truncation strategy as all the pre-trained models utilized in our study were not subjected to data filtering during their pre-training phase. Introducing significant changes to the input data structure may compromise the model's performance, as the altered input distribution might not align with the original data distribution the model was trained on.


\subsection{Feature Extraction}

During the feature extraction stage, we utilize PTMs as feature extractors to obtain representations for Title, Descripition, and Code components of a post. Different components convey information at varying degrees of detail and follow different text distributions. In our work, we consider Title, Description, and Code as three independent components and then leverage three PTMs to generate representations for each component. We then concatenate the representations of the three components as post-representation. 

\begin{figure}
	\centering
	\includegraphics[width=0.95\columnwidth]{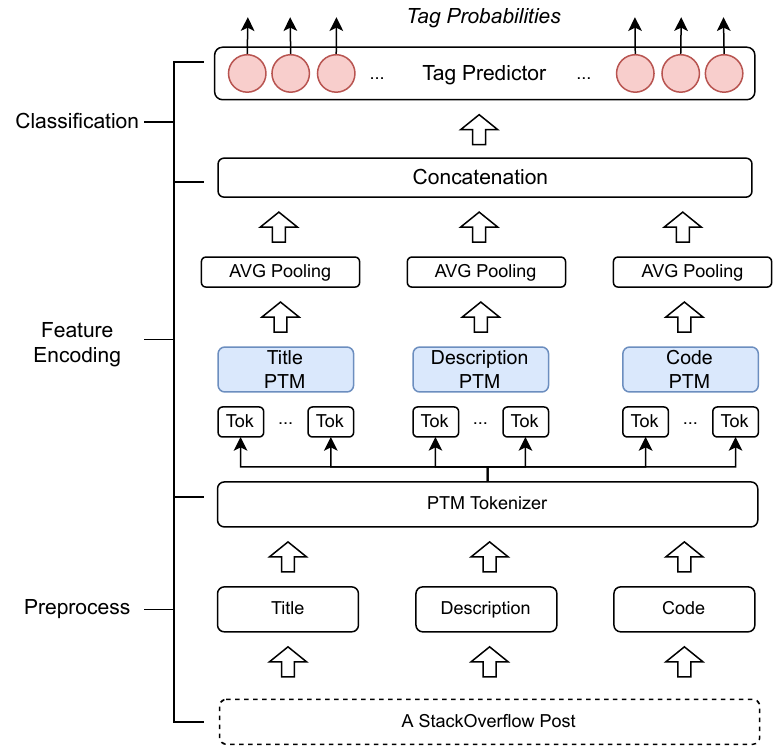}
	\caption{The overview of the \toolname framework. The title, description, and code are extracted from an SO post and fed into three different pre-trained models to obtain embeddings for each of them. A classification model takes the processed embeddings as input and produces probabilities for each tag.}
	\label{fig:architecture}
\end{figure}

\subsubsection{Language Modeling with PTMs}
We use PTMs to generate embeddings for each component. 
Training transformer-based models from scratch can be extremely resource-intensive in terms of computational cost, time, and data requirements. Without adequate pre-training, the performance of such models can be considerably suboptimal. Taking the expensive cost at the pre-training stage into account, we leverage the released PTMs by the community in the design of \toolname to generate contextual word embeddings.

The pre-trained model part of \toolname is replaceable, and we have empirically implemented eight variants of \toolname with different PTMs (refer to Section~\ref{sec:implementation}) and investigated the impact of the PTM selection within \toolnamenospace.  

\subsubsection{Pooling and Concatenation}
After the word embeddings are generated, we obtain the post representation by applying a pooling strategy.
Pooling refers to the process of transforming a sequence of (token-level) word embeddings into sentence embedding (wherein a transformer-based model, each word embedding usually has 768 hidden dimensions). A pooling strategy executes the down-sampling on the input representation, and it condenses the granular (token-level) word embeddings into a fixed-length sentence embedding that is intended to capture the meaning of the whole context.

We consider two categories of PTMs in our work, i.e., BERT-based PTMs and encoder-decoder PTMs. We discuss how to get embeddings from these two types of PTMs respectively. 
Prior studies showed that there are several common choices to derive fixed-size sentence embeddings from BERT-based PTMs~\citep{reimers2019sentencebert}, including (1) using the first \textit{CLS} token, (2) \textit{Average Pooling} and (3) \textit{Maximum Pooling}.
More specifically, Reimers and Gurevych~\citep{reimers2019sentencebert} have evaluated the effectiveness of different pooling strategies on the SNLI dataset~\citep{bowman-etal-2015-large} and the Multi-Genre NLI dataset~\citep{williams-etal-2018-broad} in the sentence classification task, and the reported \textit{Average Pooling} gives the best performance in both datasets. 
Inspired by the findings, our proposed method leverages the \textit{Average Pooling} strategy on the hidden output to generate component-wise feature vectors by default.

While BERT-based PTMs are encoder-only, Encoder-decoder PTMs have the full transformer architecture with both encoder and decoder modules. According to Ni et al.~\citep{sentencet5}, typically, strategies to obtain sentence representations for encoder-decoder PTMs are (1) \textit{Encoder-only first}: use the first token of the encoder output as the sentence embedding (2) \textit{Encoder-only mean} use the average of the encoder outputs (3) \textit{Encoder-Decoder first}: use the first decoder output as the sentence embedding. Experiments from Ni et al. showed that the \textit{Encoder-only mean} produced the best performance. Hence, we adopt the second strategy to obtain component-wise embeddings for encode-decoder PTMs.

Finally, we concatenate the three component-wise embeddings (i.e., Title, Description, and Code) sequentially to obtain the final representation of an SO post. 

\subsection{Model Training and Inference}
After performing average pooling, we concatenate the output embeddings and feed it into a feed-forward neural network to perform the task of multi-label classification.
Given a training dataset consisting of $\mathcal{X}$ (a set of SO posts) and corresponding ground truth tags $y$ for each $x \in \mathcal{X}$, we train a tag recommendation model $f$ by minimizing the following objective:

\begin{equation}
 \mathcal{L} =  - \frac{1}{N} \sum_{i=1}^{N} \sum_{j=1}^{L} y_j \times
log(f(y_j|x_i)) + (1-y_j) \times log(1-f(y_j|x_i))
\end{equation}

In the above equation, $N=|\mathcal{X}|$ is the total number of training examples and $f(y_j|x_i)$ is the probability that tag $y_j$ is related to the SO post $x_i$. The objective captures the binary cross-entropy loss on all the training examples and can be optimized by gradient descent via back-propagation. Note that the gradient flow passes through both the multi-label classifier and the PTMs used to process \textit{Title}, \textit{Description}, and \textit{Code}. The parameters of both the tag predictor and the PTMs are updated during the training process of \toolnamenospace.

Given an input $x_i$, \toolname produces a vector corresponding to all the tags. An element $f(y_j|x_i)$ in the vector corresponds to the probability that tag $y_j$ is relevant with SO post $x_i$. Stack Overflow sets a limit $k$ for the number of tags a post can have, and we rank the tags in descending order according to their probabilities produced by \toolnamenospace. The top $k$ tags with the highest probabilities are recommended for a SO post.
\section{Experimental Settings}
\label{sec:experiment}

This section introduces the research questions, describes the dataset in our experiment, the commonly-used evaluation metrics of a tag recommendation technique, and the implementation details of all considered models.

\subsection{Research Questions}

In this work, we propose \toolnamenospace, a framework that trains a transformer-based multi-label classifier to recommend tags for SO posts. To examine the effectiveness and contain a comprehensive understanding of \toolnamenospace, we are interested in answering the following four research questions:

\subsubsection*{\textbf{RQ1: Out of the eight variants of \toolname with different PTMs, which gives the best performance?}}

Transformer-based pre-trained language models have witnessed great success across multiple SE-related tasks. To the best of our knowledge, our work is the first that leverages PTMs in recommending tags for SQA sites. Past studies have shown that different PTMs have their own strengths and weaknesses. The effectiveness of PTMs varies by task due to the fact that they are pre-trained with various datasets, pre-training objectives, and vocabularies. 

For example, Mosel et al.~\citep{sebert} found BERTOverflow outperforms the NLP-domain PTM, BERT, by a substantial margin in issue type prediction and commit intent prediction. Mosel et al. further reported that the vocabulary of SE-domain PTMs (i.e., BERTOverflow) contain many programming-related vocabularies such as \textit{jvm, bugzilla, and debug} which are absent in the vocabulary of BERT \citep{bert}. However, Yang et al. reported that the NLP-domain PTMs (e.g. BERT and RoBERTa) perform better than SE-domain models, i.e., BERTOverflow in the API review classification task \citep{chengran2022saner}. Yang et al. claimed that this phenomenon may be because of the fact that the NLP-domain models are likely to be pre-trained on more extensive data. Thus, the efficacy of different pre-trained models on this task remains unclear. Moreover, in addition to the BERT-based PTMs which are encode-only, the encoder-decoder PTMs (e.g. CodeT5, PLBART, and CoText) have also performed surprisingly well in a wide spectrum of SE-related text understanding and generation tasks. It is intuitive to assume that the Encoder stacks of these models are capable of generating powerful text representation. However, not much SE literature has investigated the performance
in classification tasks. 

Since the underlying PTMs of \toolname are replaceable, it evokes our interest in investigating the effectiveness of different PTMs under the \toolname architecture and finding the most suitable PTM for SO posts tag recommendation.
Namely, we compare the results of NLP-domain BERT-based models (i.e., BERT~\citep{bert}, RoBERTa~\citep{roberta}, ALBERT~\citep{albert}), SE-domain BERT-based models (i.e., \citep{CodeBERT} and BERTOverflow~\citep{bertoverflow}), and SE-domain Seq2Seq models (i.e., PLBART~\citep{plbart}, CoTexT~\citep{cotext}, and CodeT5~\citep{codet5}).

\subsubsection*{\textbf{RQ2: How is the performance of \toolname compared to the state-of-the-art approach in Stack Overflow tag recommendation?}}

The current state-of-the-art approach for recommending tags of SO posts is implemented based on a Convolutional Neural Network and trained from scratch~\citep{post2vec}. However, Transformer-based PTMs are strengthened with the self-attention mechanism and transferred pre-train knowledge. In this research question, we investigate whether the variants of \toolname can achieve better performance than the current state-of-the-art approach.


\subsubsection*{\textbf{RQ3: Which component of post benefits \toolname the most?}}

\toolname is designed with a triplet architecture where each component of a post, i.e., Title, Description, and Code, are modeled by utilizing separate PTMs. 
Title, Description, and Code snippets complement each other and describe the post from their own perspective. Title summarizes the question with a few words; Description further expands the content from the Title; Code snippets often are a real example of the problem. Thus it motivates us to investigate which component produces the most critical contribution in the \toolname framework. 



\subsubsection*{\textbf{RQ4: How is the performance of \toolname with smaller PTMs?}}

The PTMs considered in RQ1 typically have 12 hidden layers with more than 100 million parameters. As our \toolname uses three PTMs to model different components of an SO post, the tool size is further increased by three times. Despite the encouraging performance, such a design has amplified the limitation of \toolname with respect to the inference latency. In the context of tagging SO posts, a faster inference speed could notably increase user satisfaction when running our tool. 
Therefore, it prompts our interest in experimenting with smaller PTMs under the \toolname framework. We adopted four additional smaller off-the-shelf PTMs. As introduced in Section \ref{sec:background}, they are \textit{DistilBERT}, \textit{DistilRoBERTa}, \textit{CodeBERT-small}, and \textit{CodeT5-small}. To the best of our knowledge, we leveraged all available small variants of the PTMs from RQ1.

\subsection{Data Preparation}
\label{subsec:data_preparatin}
To ensure a fair comparison to the current state-of-the-art approach~\citep{post2vec}, we select the same dataset as Xu et al. as the benchmark.
The original data is retrieved from the snapshot of the Stack Overflow dump versioned on September 5, 2018.\footnote{\url{https://archive.org/details/stackexchange}}
A tag is regarded as \textit{rare} if its occurrence is less than a pre-defined threshold $\theta$. The intuition is that if a tag appears very infrequently in such a large corpus of posts (over 11 million posts in total), it is likely to be an incorrectly created tag that developers do not broadly recognize. Therefore, we remove such rare tags as they are less important and less useful to serve as representative tags to be recommended to users~\citep{xiaxin2013}, which is a common practice in prior research~\citep{post2vec, xiaxin2013, tagcnn}.

Following the same procedure as Xu et al., we set the threshold $\theta$ for deciding a rare tag as 50. We calculate the statistics for the occurrences of tags in the dataset. While the total occurrence of all tags in our dataset is 64,197,938, the sum of occurrence for tags occurring fewer than 50 times is 33,416. This represents just 0.05\% of the total occurrences. Such a negligible percentage underlines the rarity and insignificance of these tags (tags that occur fewer than 50 times) within the entire dataset.

We remove all the rare tags of a post and the posts that contain rare tags only from the dataset. In the end, we have identified 29,357 rare tags and 23,687 common tags in total, which is the same number in Xu et al.'s work.
As a result, we obtained a dataset consisting of 10,379,014 posts. We selected 100,000 latest posts as the test data and used the rest of the 10,279,014 posts as the training data. Instead of random sampling, a chronological approach to splitting data is more representative of mimicking how the system works in real-world scenarios, especially for the Stack Overflow site. 

\subsection{Evaluation Metrics}
\label{subsec:metrics}
Previous studies of tag recommendation~\citep{post2vec, tagdc,tagcnn} on SQA sites use $Precision@k$, $Recall@k$, $F1$-$score@k$ for evaluating the performance of the approaches. Following prior studies, given a corpus of SO posts, $\mathcal{X} = \{x_1,...x_n\}$, we report $Precision@k_i$, $Recall@k_i$, $F1$-$score@k_i$ on each post $x_i$ respectively where $0 \leq i \leq n$ and calculate the average of $Precision@k_i$, $Recall@k_i$, $F1$-$score@k_i$ as $Precision@k$, $Recall@k$, $F1$-$score@k$ to be the final measure.

\vspace{2mm}

\noindent\textbf{Precision@k}
measures the average ratio of predicted ground truth tags among the list of the top-k recommended tags. For the $i$th post in the test dataset, we denote its ground truth tags for a particular post by $GT_{i}$ and predicted top-k tags of the model by $Tag_{i}^{k}$. We calculate $Precision@k_i$ as: 

\vspace{1mm}
\begin{equation}
 Precision@k_i =
\frac{  \lvert  GT_{i} \cap Tag_{i}^{k} \rvert }{k}  
\end{equation}
\vspace{1mm}

\noindent Then we average all the values of $Precision@k_i$:

\vspace{1mm}
\begin{equation}
Precision@k = \frac{ \sum_{i=1 }^{\lvert \mathcal{X}\rvert  }Precision@k_i }{ \lvert \mathcal{X} \rvert }
\end{equation}
\vspace{1mm}

\noindent\textbf{Recall@k} reports the proportion of correctly predicted ground truth tags found in the list of ground truth tags. The original formula of $Recall@k_i$ has a notable drawback: the $Recall$ score would be capped to be small when the value of $k$ is smaller than the number of ground truth tags. In the past literature~\citep{post2vec,tagcnn,tagdc}, a modified version of $Recall@k$ is commonly adopted as indicated in Equations \ref{eq:recall1} and \ref{eq:recall2}. We have adopted the modified $Recall@k$ in our work, which is as same as the one used to evaluate the current state-of-the-art approach in ~\citep{post2vec}.

\vspace{1mm}
\begin{equation}
 \label{eq:recall1}
Recall@k_i =   \begin{cases}
      \frac{| GT_{i} \cap Tag_{i}^{k}}  {k}| & \text{if } |GT_{i}| > k\\
      \frac{| GT_{i} \cap Tag_{i}^{k} | }{|GT_{i}|} & \text{if } |GT_{i}| \leq k\\
    \end{cases} \\
\end{equation}
\vspace{1mm}

\begin{equation}
\label{eq:recall2}
Recall@k=  \frac{ \sum_{i=1 }^{|\mathcal{X}|}Recall@k_i}{ | \mathcal{X} |}
\end{equation}
\vspace{1mm}

\noindent\textbf{F1-score@k} is the harmonic mean of $Precision@k$ and $Recall@k$ and it is usually considered as a summary metric. It is formally defined as:

\vspace{1mm}
\begin{equation}  \label{eq:f11}
F1\text{-}score@k_i=  2 \times \frac{ Precision@k_i \times Recall@k_i}{ Precision@k_i + Recall@k_i}
\end{equation}

\vspace{1mm}
\begin{equation}  \label{eq:f12}
F1\text{-}score@k=  \frac{ \sum_{i=1 }^{|S|}F1\text{-}score@k_i}{ | \mathcal{X} |}
\end{equation}
\vspace{1mm}

A large volume of literature on tag recommendation of SQA sites evaluates the result with k equals to 5, and 10 \citep{tagdc}. However, the number of the tags for Stack Overflow post is not allowed to be greater than 5; thus, we set the maximum value of k to 5, and we evaluate $k$ on a set of values such that $k \in \{1,2,3,4,5\}$. 

For example, assume we have a Stack Overflow post with ground truth tags (i.e., python, machine-learning, neural-network, tensorflow, keras) and a set of predicted tags (i.e., python, pytorch, neural-network, tensorflow, and keras). We calculate the Precision@5 = 0.8 (Equation 1) and Recall@5 = 0.8 (Equation 3). Thus we can use these values to compute the F1-score@5 = 0.8 (Equation 5).

\subsection{Implementation}\label{sec:implementation}
To answer the four research questions mentioned in Section \ref{sec:intro}, we train fifteen variants of \toolnamenospace. Details about each variant model are summarized in Table \ref{tab:variants}.

For RQ1, we train eight variants of \toolname by using different PTMs (i.e., CodeBERT$_{ALL}$, ALBERT$_{ALL}$, BERT$_{ALL}$, RoBERTa$_{ALL}$, BERTOverflow$_{ALL}$, CodeT5$_{ALL}$, PLBART$_{ALL}$, CoTexT$_{ALL}$ in Table~\ref{tab:variants}) and empirically investigate their performance. Each variant follows the triplet architecture (as illustrated in Figure \ref{fig:architecture}).

For RQ2, we compare the performance of \toolname (with the best performing PTM) with the state-of-the-art approach, namely Post2Vec. To reproduce the baseline introduced in Section~\ref{sec:post2vec}, we reuse the replication package\footnote{\url{https://github.com/maxxbw54/Post2Vec}} released by the original authors.

In RQ3 we develop three ablated models, CodeT5$_{NoTitle}$, CodeT5$_{NoDesc}$, and CodeT5$_{NoCode}$, (as shown in Table~\ref{tab:variants}) as CodeT5$_{ALL}$ has the best performance from the experimental results of RQ1. Different to the variants from RQ1, each ablated model only contains two components and is implemented with a Twin architecture. In another words, two PTMs are leveraged in generating post embeddings, whereas the original design of \toolname involves three PTMs. The rest of the design is much similar, where we concatenate the component representation obtained from each encoder and train a multi-label classifier. 

For RQ4, we additionally developed 4 variants of \toolname by using smaller PTMs (i.e., DistilBERT$_{ALL}$, DistilRoBERTa$_{ALL}$, CodeBERT-small$_{ALL}$, CodeT5-small$_{ALL}$ from Table~\ref{tab:variants}). We compared the performance and inference latency of these models with the best-performing variant of \toolname from RQ1. 




\begin{table}[h!]
\centering
\caption{Variants of \toolnamenospace}
\label{tab:variants}
\begin{tabular}{cccc}
\hline
\textbf{Model Name} & \textbf{BERT-Base Model} & \textbf{Considered Components} & \textbf{Architecture} \\ \hline
BERT$_{ALL}$           & BERT           & Title, Description,Code & Triplet \\
RoBERTa$_{ALL}$        & RoBERTa        & Title, Description,Code & Triplet \\
ALBERT$_{ALL}$         & ALBERT         & Title, Description,Code & Triplet \\
CodeBERT$_{ALL}$       & CodeBERT       & Title, Description,Code & Triplet \\
BERTOverflow$_{ALL}$   & BERTOverflow   & Title, Description,Code & Triplet \\
CodeT5$_{ALL}$         & CodeT5         & Title, Description,Code & Triplet \\
PLBART$_{ALL}$         & PLBART         & Title, Description,Code & Triplet \\
CoTexT$_{ALL}$         & CoText         & Title, Description,Code & Triplet \\
CodeT5$_{NoTitle}$     & CodeT5         & Description,Code        & Twin    \\
CodeT5$_{NoDesc}$      & CodeT5         & Title, Code             & Twin    \\
CodeT5$_{NoCode}$      & CodeT5         & Title, Description      & Twin    \\
DistilBERT$_{ALL}$     & DistilBERT     & Title, Description,Code & Triplet \\
DistilRoBERTa$_{ALL}$  & DistilRoBERTa  & Title, Description,Code & Triplet \\
CodeBERT-small$_{ALL}$ & CodeBERT-small & Title, Description,Code & Triplet \\
CodeT5-small$_{ALL}$   & CodeT5-small   & Title, Description,Code & Triplet \\ \hline
\end{tabular}
\end{table}

All the variants of \toolname are implemented with PyTorch V.1.10.0\footnote{\url{https://pytorch.org}} and HuggingFace Transformer library V.4.12.3\footnote{\url{https://huggingface.co}}. Considering the extensive amount of the data set, we only trained the models for one epoch at the fine-tuning stage. For each variant, we set the batch size as 64. We set the initial learning rate as 7E-05 and applied a linear scheduler to control the learning rate at run time.

\section{Experimental Results}
\label{sec:results}
In this section, we report the experimental results of the variants under our proposed framework and the baseline approach. We further conduct an ablation study on our best-performing variant to assess the importance of the underlying component. 
Finally, we report the performance of \toolname on smaller PTMs with reduced model size. 
Based on the results, we answer the research questions presented in Section~\ref{sec:experiment}.

\subsection*{ \textbf{RQ1. Out of the eight variants of \toolname with different PTMs, which gives the best performance?}}
\begin{table}[]
\centering
\caption{Comparison of all variants of \toolname with a triplet architecture and the baseline approach Post2Vec.}
\label{tab:all_results}
\resizebox{\textwidth}{!}{%
\begin{tabular}{c|c|c|ccccc}
\hline
\multirow{2}{*}{\textbf{Architecture}}   & \multirow{2}{*}{\textbf{Domain}} & \multirow{2}{*}{\textbf{Model Name}}      & \multicolumn{5}{c}{\textbf{Precision@k}}                                                                                                                               \\ \cline{4-8} 
                                 &                                  &                                           & \multicolumn{1}{c|}{\textbf{P@1}}   & \multicolumn{1}{c|}{\textbf{P@2}}   & \multicolumn{1}{c|}{\textbf{P@3}}   & \multicolumn{1}{c|}{\textbf{P@4}}   & \textbf{P@5}   \\ \hline
\multirow{5}{*}{Encoder-only}    & \multirow{2}{*}{SE}              & CodeBERT$_{ALL}$                          & \multicolumn{1}{c|}{0.848}          & \multicolumn{1}{c|}{0.701}          & \multicolumn{1}{c|}{0.579}          & \multicolumn{1}{c|}{0.486}          & 0.415          \\ \cline{3-8} 
                                 &                                  & BERTOverflow$_{ALL}$                      & \multicolumn{1}{c|}{0.725}          & \multicolumn{1}{c|}{0.592}          & \multicolumn{1}{c|}{0.489}          & \multicolumn{1}{c|}{0.412}          & 0.354          \\ \cline{2-8} 
                                 & \multirow{3}{*}{NLP}             & RoBERTa$_{ALL}$                           & \multicolumn{1}{c|}{0.843}          & \multicolumn{1}{c|}{0.694}          & \multicolumn{1}{c|}{0.571}          & \multicolumn{1}{c|}{0.478}          & 0.409          \\ \cline{3-8} 
                                 &                                  & BERT$_{ALL}$                              & \multicolumn{1}{c|}{0.845}          & \multicolumn{1}{c|}{0.696}          & \multicolumn{1}{c|}{0.575}          & \multicolumn{1}{c|}{0.482}          & 0.413          \\ \cline{3-8} 
                                 &                                  & ALBERT$_{ALL}$                            & \multicolumn{1}{c|}{0.748}          & \multicolumn{1}{c|}{0.586}          & \multicolumn{1}{c|}{0.469}          & \multicolumn{1}{c|}{0.386}          & 0.327          \\ \hline
\multirow{3}{*}{Encoder-Decoder} & \multirow{3}{*}{SE}              & \textbf{CodeT5$_{ALL}$}                   & \multicolumn{1}{c|}{\textbf{0.855}} & \multicolumn{1}{c|}{\textbf{0.708}} & \multicolumn{1}{c|}{\textbf{0.586}} & \multicolumn{1}{c|}{\textbf{0.492}} & \textbf{0.420} \\ \cline{3-8} 
                                 &                                  & PLBART$_{ALL}$                            & \multicolumn{1}{c|}{0.821}          & \multicolumn{1}{c|}{0.669}          & \multicolumn{1}{c|}{0.547}          & \multicolumn{1}{c|}{0.456}          & 0.388          \\ \cline{3-8} 
                                 &                                  & CoTexT$_{ALL}$                            & \multicolumn{1}{c|}{0.848}          & \multicolumn{1}{c|}{0.701}          & \multicolumn{1}{c|}{0.579}          & \multicolumn{1}{c|}{0.486}          & 0.415          \\ \hline
-                                & -                                & Post2Vec                                  & \multicolumn{1}{c|}{0.786}          & \multicolumn{1}{c|}{0.628}          & \multicolumn{1}{c|}{0.507}          & \multicolumn{1}{c|}{0.421}          & 0.359          \\ \hline
\multirow{2}{*}{\textbf{Architecture}}   & \multirow{2}{*}{\textbf{Domain}} & \multirow{2}{*}{\textbf{Model Name}}      & \multicolumn{5}{c}{\textbf{Recall@k}}                                                                                                                                  \\ \cline{4-8} 
                                 &                                  &                                           & \multicolumn{1}{c|}{\textbf{R@1}}   & \multicolumn{1}{c|}{\textbf{R@2}}   & \multicolumn{1}{c|}{\textbf{R@3}}   & \multicolumn{1}{c|}{\textbf{R@4}}   & \textbf{R@5}   \\ \hline
\multirow{5}{*}{Encoder-only}    & \multirow{2}{*}{SE}              & CodeBERT$_{ALL}$                          & \multicolumn{1}{c|}{0.848}          & \multicolumn{1}{c|}{0.756}          & \multicolumn{1}{c|}{0.724}          & \multicolumn{1}{c|}{0.733}          & 0.757          \\ \cline{3-8} 
                                 &                                  & \multicolumn{1}{l|}{BERTOverflow$_{ALL}$} & \multicolumn{1}{c|}{0.725}          & \multicolumn{1}{c|}{0.635}          & \multicolumn{1}{c|}{0.607}          & \multicolumn{1}{c|}{0.619}          & 0.644          \\ \cline{2-8} 
                                 & \multirow{3}{*}{NLP}             & RoBERTa$_{ALL}$                           & \multicolumn{1}{c|}{0.843}          & \multicolumn{1}{c|}{0.747}          & \multicolumn{1}{c|}{0.714}          & \multicolumn{1}{c|}{0.722}          & 0.746          \\ \cline{3-8} 
                                 &                                  & BERT$_{ALL}$                              & \multicolumn{1}{c|}{0.845}          & \multicolumn{1}{c|}{0.750}          & \multicolumn{1}{c|}{0.719}          & \multicolumn{1}{c|}{0.728}          & 0.752          \\ \cline{3-8} 
                                 &                                  & ALBERT$_{ALL}$                            & \multicolumn{1}{c|}{0.748}          & \multicolumn{1}{c|}{0.630}          & \multicolumn{1}{c|}{0.588}          & \multicolumn{1}{c|}{0.588}          & 0.605          \\ \hline
\multirow{3}{*}{Encoder-Decoder} & \multirow{3}{*}{SE}              & \textbf{CodeT5$_{ALL}$}                   & \multicolumn{1}{c|}{\textbf{0.855}} & \multicolumn{1}{c|}{\textbf{0.763}} & \multicolumn{1}{c|}{\textbf{0.732}} & \multicolumn{1}{c|}{\textbf{0.741}} & \textbf{0.765} \\ \cline{3-8} 
                                 &                                  & PLBART$_{ALL}$                            & \multicolumn{1}{c|}{0.821}          & \multicolumn{1}{c|}{0.720}          & \multicolumn{1}{c|}{0.683}          & \multicolumn{1}{c|}{0.688}          & 0.709          \\ \cline{3-8} 
                                 &                                  & CoTexT$_{ALL}$                            & \multicolumn{1}{c|}{0.848}          & \multicolumn{1}{c|}{0.755}          & \multicolumn{1}{c|}{0.723}          & \multicolumn{1}{c|}{0.733}          & 0.756          \\ \hline
-                                & -                                & Post2Vec                                  & \multicolumn{1}{c|}{0.786}          & \multicolumn{1}{c|}{0.678}          & \multicolumn{1}{c|}{0.636}          & \multicolumn{1}{c|}{0.639}          & 0.659          \\ \hline
\multirow{2}{*}{\textbf{Architecture}}   & \multirow{2}{*}{\textbf{Domain}} & \multirow{2}{*}{\textbf{Model Name}}      & \multicolumn{5}{c}{\textbf{F1-score@k}}                                                                                                                                \\ \cline{4-8} 
                                 &                                  &                                           & \multicolumn{1}{c|}{\textbf{F@1}}   & \multicolumn{1}{c|}{\textbf{F@2}}   & \multicolumn{1}{c|}{\textbf{F@3}}   & \multicolumn{1}{c|}{\textbf{F@4}}   & \textbf{F@5}   \\ \hline
\multirow{5}{*}{Encoder-only}    & \multirow{2}{*}{SE}              & CodeBERT$_{ALL}$                          & \multicolumn{1}{c|}{0.848}          & \multicolumn{1}{c|}{0.719}          & \multicolumn{1}{c|}{0.625}          & \multicolumn{1}{c|}{0.561}          & 0.513          \\ \cline{3-8} 
                                 &                                  & \multicolumn{1}{l|}{BERTOverflow$_{ALL}$} & \multicolumn{1}{c|}{0.725}          & \multicolumn{1}{c|}{0.606}          & \multicolumn{1}{c|}{0.527}          & \multicolumn{1}{c|}{0.475}          & 0.427          \\ \cline{2-8} 
                                 & \multirow{3}{*}{NLP}             & RoBERTa$_{ALL}$                           & \multicolumn{1}{c|}{0.843}          & \multicolumn{1}{c|}{0.711}          & \multicolumn{1}{c|}{0.617}          & \multicolumn{1}{c|}{0.553}          & 0.505          \\ \cline{3-8} 
                                 &                                  & BERT$_{ALL}$                              & \multicolumn{1}{c|}{0.845}          & \multicolumn{1}{c|}{0.714}          & \multicolumn{1}{c|}{0.621}          & \multicolumn{1}{c|}{0.557}          & 0.510          \\ \cline{3-8} 
                                 &                                  & ALBERT$_{ALL}$                            & \multicolumn{1}{c|}{0.748}          & \multicolumn{1}{c|}{0.600}          & \multicolumn{1}{c|}{0.506}          & \multicolumn{1}{c|}{0.447}          & 0.406          \\ \hline
\multirow{3}{*}{Encoder-Decoder} & \multirow{3}{*}{SE}              & \textbf{CodeT5$_{ALL}$}                   & \multicolumn{1}{c|}{\textbf{0.855}} & \multicolumn{1}{c|}{\textbf{0.726}} & \multicolumn{1}{c|}{\textbf{0.633}} & \multicolumn{1}{c|}{\textbf{0.568}} & \textbf{0.519} \\ \cline{3-8} 
                                 &                                  & PLBART$_{ALL}$                            & \multicolumn{1}{c|}{0.821}          & \multicolumn{1}{c|}{0.686}          & \multicolumn{1}{c|}{0.590}          & \multicolumn{1}{c|}{0.526}          & 0.480          \\ \cline{3-8} 
                                 &                                  & CoTexT$_{ALL}$                            & \multicolumn{1}{c|}{0.848}          & \multicolumn{1}{c|}{0.719}          & \multicolumn{1}{c|}{0.625}          & \multicolumn{1}{c|}{0.561}          & 0.513          \\ \hline
-                                & -                                & Post2Vec                                  & \multicolumn{1}{c|}{0.786}          & \multicolumn{1}{c|}{0.646}          & \multicolumn{1}{c|}{0.549}          & \multicolumn{1}{c|}{0.488}          & 0.445          \\ \hline
\end{tabular}%
}
\end{table}

\begin{figure}
\centering
	\includegraphics[width=0.75\linewidth]{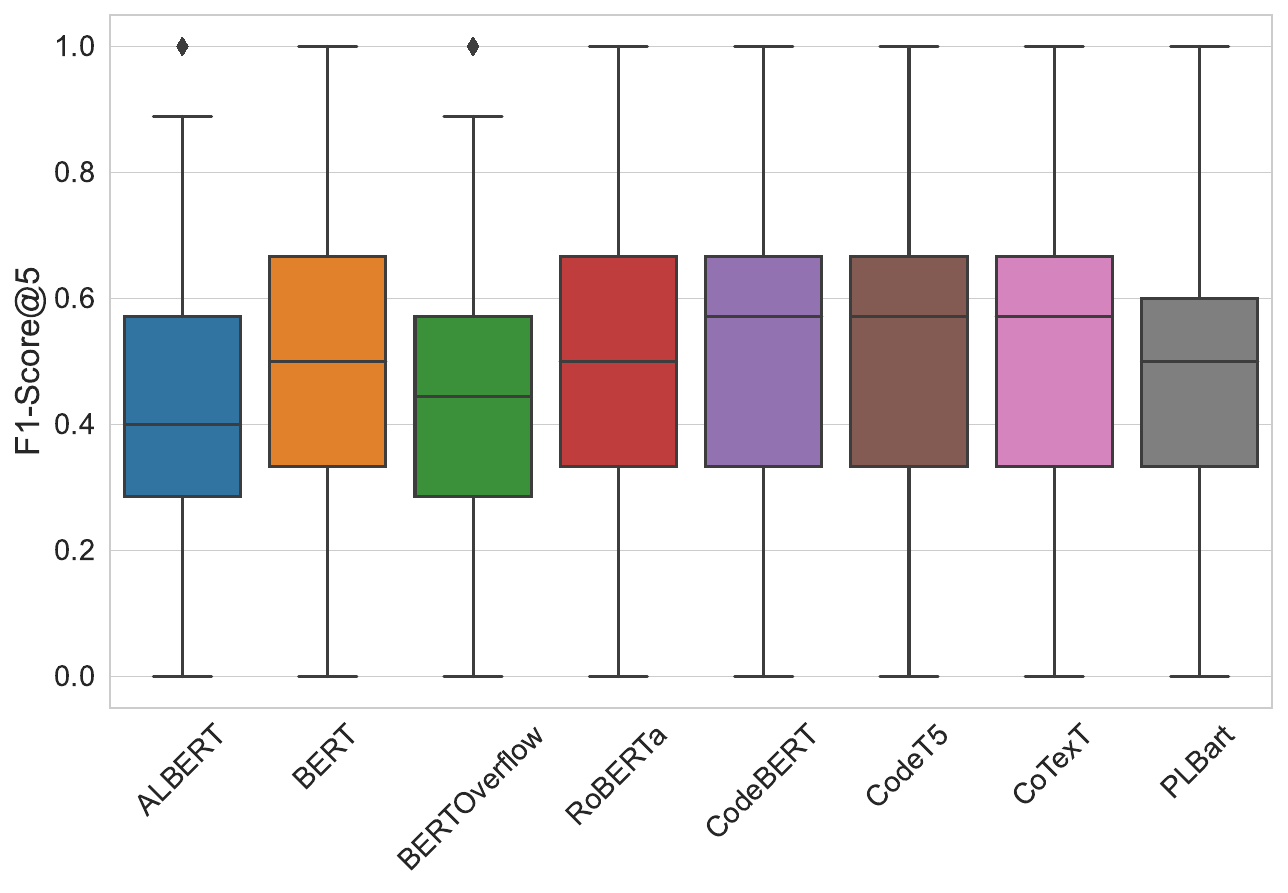}
	\vspace{-2mm}
	\caption{Distribution of F1-Score@5 for All \toolname Variants Using a Triplet Architecture }
	\label{fig:rq1-f1@5}
	\vspace{-4mm}
\end{figure}

\noindent\textbf{Results and Analysis} 
To answer the question, we report the performance on eight variants of \toolname. We leverage three NLP-domain BERT-based PTMs (BERT, RoBERTa, and ALBERT), two SE-domain BERT-based PTMs (CodeBERT and BERTOverflow), and three SE-domain encoder-decoder PTMs (CodeT5, PLBART, and CoTexT). 
These variants are implemented with the Triplet architecture, which considers Title, Description, and Code as input. We refer to these variant models as BERT$_{ALL}$, RoBERTa$_{ALL}$, ALBERT$_{ALL}$, CodeBERT$_{ALL}$, BERTOverflow$_{ALL}$, CodeT5$_{ALL}$, PLBART$_{ALL}$, and CoText$_{ALL}$ respectively. 

Table \ref{tab:all_results} illustrates the results of applying different PTMs in recommending the tags of SO posts and Figure \ref{fig:rq1-f1@5} draws the boxplot on the distribution of $F1$-$score@5$ for each variant. 
Performance-wise, CodeT5$_{ALL}$ achieves the highest rank and consistently outperforms the other variants in all evaluation metrics. For $F1$-$score@1$-$5$, it obtains the performance of 0.855, 0.726, 0.633, 0.568, and 0.519, respectively. ALBERT is the worst-performing model and BERTOverflow$_{ALL}$ is merely better than ALBERT$_{ALL}$. 

Overall, BERTOverflow$_{ALL}$ and ALBERT$_{ALL}$ only achieve 0.427 and 0.406 in $F1$-$score@5$, which are lower than the other variants by a substantial margin. CodeBERT$_{ALL}$ and CoTexT$_{ALL}$ are tied for the second-best performing models, which both obtain an $F1$-$score@5$ of 0.513. In the next line, BERT$_{ALL}$ and RoBERTa$_{ALL}$ loseCodeBERT$_{ALL}$ and CoTexT$_{ALL}$ only by a small margin.

From Figure \ref{fig:rq1-f1@5}, we can see that the 25\% percentile value across most model are around 0.333 for $F1$-$score@5$. The median (50\% percentile) value is a key metric for central tendency, and most models center around the 0.5 mark. However, CodeBERT$_{ALL}$, CoTexT$_{ALL}$, and CodeT5$_{ALL}$ slightly surpass this general trend, showcasing a median of more than 0.571.


As CodeT5$_{ALL}$, CoTexT$_{ALL}$ , and PLBART$_{ALL}$ all achieve promising results, we demonstrate that encoder-decoder PTMs are also capable of generating good representations for SO posts. A possible reason for it is that these models adopt multi-task learning objectives at the training stage. 
Different to BERT-based PTMs, which follow the \textit{pretrain-then-finetune} paradigm, encoder-decoder models like CodeT5 are trained with multiple tasks and multiple datasets at the same time (multi-task learning). 
Previous literature has claimed that multi-task learning leads to more generalized and better representations when being adapted to new tasks and domains ~\citep{liu2019multi}, our results conform to this claim. 
Moreover, CodeT5 uses more training data than CodeBERT and has more pre-training tasks than CoTexT. These may be attributed to its outstanding performance.

ALBERT$_{ALL}$ and BERTOverflow$_{ALL}$ are largely outperformed by the other variants. One possible reason for this could be that ALBERT has fewer parameters since its design aspires to address the GPU memory limitation. BERTOverflow follows the same architecture as BERT, and is designed with a vocabulary better suited to the software engineering domain~\citep{sebert}, but it performs much worse than BERT$_{ALL}$ and RoBERTa$_{ALL}$ by a large margin. However, the experimental results suggested that BERTOverflow may still require additional training. It is potentially caused by the quality and size of the datasets used at the pre-training stage. BERTOverflow is pre-trained with 152 million sentences from SO. BERT is trained on the entire English Wikipedia and the Book Corpus dataset, written by professionals and constantly reviewed. However, sentences from SO can be written by arbitrary authors and the existence of in-line code within a post would introduce extra noise. Additionally, the training corpus of BERT contains 3.3 billion words in total, and the average sentence length of BookCorpus is 11 words. By estimation, the training corpus of BERT is likely to be twice more than BERTOverflow.

Moreover, although CodeBERT and BERTOverflow are both SE-domain PTMs, the performance of these two models is very different. This phenomenon could be because of that CodeBERT and BERTOverflow are initialized with different strategies before the pre-training starts. CodeBERT is initialized based on RoBERTa's checkpoint, whereas BERTOverflow is trained from scratch with SO data. Initializing with the checkpoint of another pre-trained model can inherit their knowledge and typically reduce the training effort by orders of magnitude~\citep{rothe-etal-2020-leveraging}. 

Comparing CodeBERT with BERT and RoBERTa, CodeBERT has utilized both natural language and programming language at the pre-training stage, while the other two models are trained solely with natural language data. As a result, CodeBERT is better at understanding programming languages and SE terminology. Given that many SO posts contain code snippets, this advantage allows CodeBERT$_{ALL}$ to achieve improved performance. In addition, the good performance of BERT and RoBERTa also suggests that pre-training models with a large scale of natural language data also be beneficial for programming language modeling.


Furthermore, conventional PTMs accept a maximum input sequence length of 512 sub-tokens. Our method utilizes a \textit{head-only} truncation strategy by default, we further experiment with the effect of the \textit{tail-only} truncation strategy. In Table \ref{tab:tail}, we present the performance of various variants of \toolnamenospace, in terms of $F1$-$score@k$ at different levels (k ranging from 1 to 5) with the tail-only truncation strategy. For convenience, we also add  $F1$-$score@k$ of head-only truncation strategy in Table \ref{tab:tail}. 
When comparing the two truncation strategies for each variant, the performance differences are minimal. CodeT5\(_{ALL}\) consistently showed the highest F1-scores across all values of \(k\) for both truncation strategies. 
Most variants display slightly better performance under the head-only truncation strategy as compared to the tail-only truncation. Notably, RoBERTa${_ALL}$ shows a more significant drop in performance with tail-only truncation, especially for F1-score@1 and F1-score@2. This suggests that the head portion of the input might be more crucial for RoBERTa$_{ALL}$. CodeBERT$_{ALL}$, BERT$_{ALL}$, CodeT5$_{ALL}$, and CoTexT$_{ALL}$ show minimal variations in their performance across the two truncation strategies.
The results presented emphasize the importance of PTM selection under the framework of \toolnamenospace, while the choice between head-only and tail-only truncation doesn't lead to vast performance differences. 
Given that the \textit{head-only} truncation method has a slightly better $F1$-$score@k$, it has been selected as the default truncation strategy for subsequent experiments.








\begin{table}[h!]
\centering
\caption{Results of Variants of \toolname for F1-score@k (k ranging from 1 to 5) with Head-only and Tail-only Truncation Strategies}
\label{tab:tail}
\begin{tabular}{c|c|c|c|c|c}
\hline
\multicolumn{6}{c}{\textbf{Head-only Truncation}} \\
\hline
\textbf{Model Name} & \textbf{F@1} & \textbf{F@2} & \textbf{F@3} & \textbf{F@4} & \textbf{F@5} \\
\hline
CodeBERT$_{ALL}$ & 0.848 & 0.719 & 0.625 & 0.561 & 0.513 \\
BERT$_{ALL}$ & 0.845 & 0.714 & 0.621 & 0.557 & 0.510 \\
RoBERTa$_{ALL}$ & 0.843 & 0.711 & 0.617 & 0.553 & 0.505 \\
BERTOverflow$_{ALL}$ & 0.725 & 0.606 & 0.527 & 0.475 & 0.427 \\
ALBERT$_{ALL}$ & 0.748 & 0.600 & 0.506 & 0.447 & 0.406 \\
CodeT5$_{ALL}$ & 0.855 & 0.726 & 0.633 & 0.568 & 0.519 \\
PLBART$_{ALL}$ & 0.821 & 0.686 & 0.590 & 0.526 & 0.480 \\
CoTexT$_{ALL}$ & 0.848 & 0.719 & 0.625 & 0.561 & 0.513 \\
\hline
\multicolumn{6}{c}{\textbf{Tail-only Truncation}} \\
\hline
\textbf{Model Name} & \textbf{F@1} & \textbf{F@2} & \textbf{F@3} & \textbf{F@4} & \textbf{F@5} \\
\hline
CodeBERT$_{ALL}$ & 0.847 & 0.718 & 0.624 & 0.560 & 0.512 \\
BERT$_{ALL}$ & 0.844 & 0.714 & 0.620 & 0.556 & 0.509 \\
RoBERTa$_{ALL}$ & 0.818 & 0.691 & 0.600 & 0.539 & 0.494 \\
BERTOverflow$_{ALL}$ & 0.722 & 0.604 & 0.524 & 0.473 & 0.425 \\
ALBERT$_{ALL}$ & 0.746 & 0.598 & 0.504 & 0.445 & 0.404 \\
CodeT5$_{ALL}$ & 0.854 & 0.725 & 0.632 & 0.567 & 0.518 \\
PLBART$_{ALL}$ & 0.821 & 0.685 & 0.590 & 0.525 & 0.479 \\
CoTexT$_{ALL}$ & 0.847 & 0.718 & 0.624 & 0.560 & 0.512 \\
\hline
\end{tabular}
\end{table}

\begin{tcolorbox}
 \textbf{Answers to RQ1}: Among the eight considered PTMs of \toolnamenospace, the one implemented with CodeT5 produces the best performance. Most PTMs from the SE domain give a more promising performance than PTMs from the NLP domain under \toolnamenospace.
\end{tcolorbox}

\subsection*{\textbf{RQ2. How is the performance of \toolname compared to the state-of-the-art approach in Stack Overflow tag recommendation?}}

\noindent\textbf{Results and Analysis}
As presented in Table~\ref{tab:all_results}, the best performing variant of \toolnamenospace, i.e., CodeT5$_{ALL}$, substantially outperforms Post2Vec. In terms of $F1$-$score@k$ (where k ranges from 1 to 5), CodeT5$_{ALL}$ improved the performance by 8.8\%, 12.4\%, 15.3\%, 16.4\%, and 16.6\%. On the other hand, CodeBERT$_{ALL}$ and CoTexT$_{ALL}$ surpass the performance of Post2Vec by 7.9\%, 11.3\%, 13.8\%, 15.0\% and 15.3\%, respectively;
BERT$_{ALL}$, RoBERTa$_{ALL}$, and PLBART$_{ALL}$ are also able to outperform Post2Vec by 4.4\% to 14.6\%. 
However, not all PTMs demonstrated exceptional performance under \toolnamenospace. Post2Vec outperformed BERTOverflow$_{ALL}$ by 8.4\%, 6.6\%, 8.5\%, 2.7\%, and 4.7\%, and outperformed ALBERT$_{ALL}$ by 5.1\%, 7.7\%, 8.5\%, 9.2\%, and 9.6\% in $F1$-$score@1$--$5$.


Different from Post2Vec, \toolname involves a vast amount of knowledge accumulated from the dataset used for pre-training. \toolname leverages PTMs to extract feature vectors and optimize the post representation during the fine-tuning stage, whereas Post2Vec learns post representations from scratch. Thus, PTMs give a better initialization of the model. Furthermore, CodeT5 provides in-domain knowledge of SE. Our results indicate that the knowledge learned in the pre-training stage is valuable to the success of the tag recommendation task. 

Another potential reason for the superior performance of \toolname is that transformer-based models are more powerful than CNN in capturing long-range dependencies~\citep{transformer}. The architecture of BERT-based PTMs is inherited from a Transformer. One of the critical concepts of Transformers is the \textit{self-attention mechanism}, which enables its ability to capture long dependencies among all input sequences.
Our results demonstrate the effectiveness and generalizability of transfer learning and reveal that the PTMs can achieve outstanding performance in the tag recommendation task for SO posts.

\begin{tcolorbox}
    \textbf{Answers to RQ2}:
    CodeT5$_{ALL}$, PLBART$_{ALL}$, CoTexT$_{ALL}$,
    CodeBERT$_{ALL}$, BERT$_{ALL}$, and RoBERTa$_{ALL}$ outperform the state-of-the-art approach by a substantial margin. However, BERTOverflow$_{ALL}$ and ALBERT$_{ALL}$ demonstrated worse performance than the state-of-the-art approach.
\end{tcolorbox}

\subsection*{\textbf{RQ3. Which component of post benefits \toolname the most?}}

 \begin{table}[ht]
\centering
\caption{Experiment Results of RQ3: Ablation study for post components using both CodeT5 and CodeBERT models.}
\label{tab:rq3_combined}
\begin{tabular}{c|ccccc}
\hline
\multirow{2}{*}{\textbf{Model Name}} & \multicolumn{5}{c}{\textbf{Precision@k}} \\  \cline{2-6} 
                                     & P@1 & P@2 & P@3 & P@4 & P@5 \\ \hline
\textbf{CodeT5$_{\textbf{ALL}}$}     & \textbf{0.855} & \textbf{0.708} & \textbf{0.586} & \textbf{0.492} & \textbf{0.420} \\
CodeT5$_{NoCode}$                    & 0.822 & 0.677 & 0.558 & 0.470 & 0.403 \\
CodeT5$_{NoDesc}$                    & 0.821 & 0.669 & 0.547 & 0.456 & 0.388 \\
CodeT5$_{NoTitle}$                   & 0.817 & 0.673 & 0.557 & 0.468 & 0.401 \\ 
\textbf{CodeBERT$_{\textbf{ALL}}$}   & \textbf{0.848} & \textbf{0.701} & \textbf{0.579} & \textbf{0.486} & \textbf{0.415} \\
CodeBERT$_{NoCode}$                  & 0.823 & 0.682 & 0.562 & 0.472 & 0.408 \\
CodeBERT$_{NoDesc}$                  & 0.822 & 0.671 & 0.549 & 0.458 & 0.390 \\
CodeBERT$_{NoTitle}$                 & 0.808 & 0.664 & 0.547 & 0.460 & 0.394 \\ \hline

\multirow{2}{*}{\textbf{Model Name}} & \multicolumn{5}{c}{\textbf{Recall@k}} \\  \cline{2-6} 
                                     & R@1 & R@2 & R@3 & R@4 & R@5 \\ \hline
\textbf{CodeT5$_{\textbf{ALL}}$}     & \textbf{0.855} & \textbf{0.763} & \textbf{0.732} & \textbf{0.741} & \textbf{0.765} \\
CodeT5$_{NoCode}$                    & 0.822 & 0.728 & 0.697 & 0.707 & 0.732 \\
CodeT5$_{NoDesc}$                    & 0.821 & 0.720 & 0.684 & 0.689 & 0.710 \\
CodeT5$_{NoTitle}$                   & 0.817 & 0.724 & 0.695 & 0.705 & 0.730 \\ 
\textbf{CodeBERT$_{\textbf{ALL}}$}   & \textbf{0.848} & \textbf{0.756} & \textbf{0.724} & \textbf{0.733} & \textbf{0.757} \\
CodeBERT$_{NoCode}$                  & 0.823 & 0.733 & 0.702 & 0.712 & 0.737 \\
CodeBERT$_{NoDesc}$                  & 0.822 & 0.723 & 0.686 & 0.693 & 0.714 \\
CodeBERT$_{NoTitle}$                 & 0.808 & 0.715 & 0.683 & 0.693 & 0.718 \\ \hline

\multirow{2}{*}{\textbf{Model Name}} & \multicolumn{5}{c}{\textbf{F1-score@k}} \\  \cline{2-6} 
                                     & F@1 & F@2 & F@3 & F@4 & F@5 \\ \hline
\textbf{CodeT5$_{\textbf{ALL}}$}     & \textbf{0.855} & \textbf{0.726} & \textbf{0.633} & \textbf{0.568} & \textbf{0.519} \\
CodeT5$_{NoCode}$                    & 0.822 & 0.694 & 0.603 & 0.543 & 0.499 \\
CodeT5$_{NoDesc}$                    & 0.821 & 0.686 & 0.591 & 0.527 & 0.480 \\
CodeT5$_{NoTitle}$                   & 0.817 & 0.690 & 0.600 & 0.541 & 0.496 \\ 
\textbf{CodeBERT$_{\textbf{ALL}}$}   & \textbf{0.848} & \textbf{0.719} & \textbf{0.625} & \textbf{0.561} & \textbf{0.513} \\ 
CodeBERT$_{NoCode}$                  & 0.823 & 0.699 & 0.607 & 0.545 & 0.500 \\
CodeBERT$_{NoDesc}$                  & 0.822 & 0.688 & 0.593 & 0.530 & 0.483 \\
CodeBERT$_{NoTitle}$                 & 0.808 & 0.680 & 0.591 & 0.531 & 0.487 \\ \hline
\end{tabular}
\end{table}

\noindent\textbf{Results and Analysis} To answer this research question, we conduct an ablation study to investigate the importance of each component, i.e., Title, Description, and Code, respectively. Note that \toolname is implemented with a triplet architecture by default. To answer this research question, we modified it to a twin architecture to fit two considered components at a time. We train three ablated models with our identified best-performing PTM, i.e., CodeT5.

\begin{figure}[ht]
    \label{fig:graph}
	\includegraphics[width=1\linewidth]{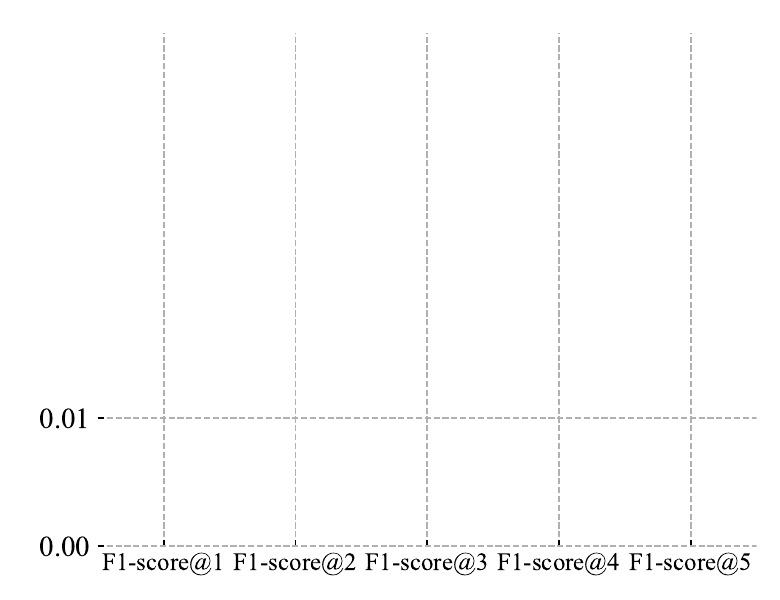}
	\caption{A line chart demonstrate performance difference in $F1$-$score@k$ between each ablated models and CodeT5$_{ALL}$, where $k \in \{1,2,3,4,5\}$. The value on the y axis is calculated using the corresponding score of the candidate ablated model minus the corresponding score of CodeT5$_{ALL}$ .}
	\label{fig:f1}
\end{figure}

The results for RQ3 are presented in Table \ref{tab:rq3_combined}. Notice that Table \ref{tab:rq3_combined} also contains the results for the ablated model for CodeBERT$_{ALL}$, which is the best-performing variant in our previous ICPC paper. From the table, we identify that CodeT5$_{ALL}$ remains to be the best-performing model on all evaluation metrics. The results of both variants CodeT5$_{ALL}$ and CodeBERT$_{ALL}$ show that code plays the least important role. 
Excluding title and description also leads to a decline in all metrics, but the drop is less severe than when removing the code.

To provide a more intuitive understanding of the result, we further illustrate the performance gap of $F1$-$score@1$--$5$ between the ablated models and CodeT5$_{ALL}$ by visualizing in Figure \ref{fig:f1}, where the value on the y axis is calculated using the score of CodeT5$_{ALL}$ minus the score of the ablated model. CodeT5$_{NoCode}$ yielded the most promising performance among the ablated models, which implies that the code snippets are beneficial, but they are the least significant among all three components.

An interesting finding is that CodeT5$_{NoDesc}$ performed better in $F1$-$score@1$ and CodeT5$_{NoTitle}$ performed better in $F1$-$score@2$--$5$. It implies that Title is more important for boosting the performance of $F1$-$score@1$ and Description is more critical for improving $F1$-$score@2$--$5$. 
A possible explanation could be that Title always succinctly describes a post's central message, which could directly help the system predict the top one tag. Description is usually much longer and elaborates the Title with more explanations; thus, it is more beneficial to recommend multiple tags. Moreover, as CodeT5$_{ALL}$ is still the best-performing model, it confirms that Code is a meaningful component and we need all three components in the tag recommendation task of SO. 

\begin{tcolorbox}
    \textbf{Answers to RQ3}: 
    Under \toolnamenospace, Title and Description are more important than Code for tag recommendation. Title plays the most important role in predicting the top-1 most relevant tags, and the contribution of Description increases when the number of predicted tags increases from two. Still, considering all three components achieve the best performance.
\end{tcolorbox}

\subsection*{\textbf{RQ4. How is the performance of \toolname with with smaller PTMs?}}

\begin{table}[h]
\centering
\caption{Comparison of variants of \toolname with smaller pre-trained models and the best-performing variant of \toolname}
\label{tab:rq4}
\begin{tabular}{c|ccccc}
\hline
\multirow{2}{*}{\textbf{Model Name}} & \multicolumn{5}{c}{\textbf{Precision@k}}                                                                                                                               \\ \cline{2-6} 
                                     & \multicolumn{1}{c|}{P@1}            & \multicolumn{1}{c|}{P@2}            & \multicolumn{1}{c|}{P@3}            & \multicolumn{1}{c|}{P@4}            & P@5            \\ \hline
\textbf{CodeT5$_{ALL}$}              & \multicolumn{1}{c|}{\textbf{0.855}} & \multicolumn{1}{c|}{\textbf{0.708}} & \multicolumn{1}{c|}{\textbf{0.586}} & \multicolumn{1}{c|}{\textbf{0.492}} & \textbf{0.420} \\ \hline
DistilBERT$_{ALL}$                   & \multicolumn{1}{c|}{0.835}          & \multicolumn{1}{c|}{0.684}          & \multicolumn{1}{c|}{0.560}          & \multicolumn{1}{c|}{0.468}          & 0.399          \\ \hline
DistilRoBERTa$_{ALL}$                & \multicolumn{1}{c|}{0.830}          & \multicolumn{1}{c|}{0.679}          & \multicolumn{1}{c|}{0.557}          & \multicolumn{1}{c|}{0.464}          & 0.396          \\ \hline
CodeBERT-small$_{ALL}$               & \multicolumn{1}{c|}{0.831}          & \multicolumn{1}{c|}{0.681}          & \multicolumn{1}{c|}{0.559}          & \multicolumn{1}{c|}{0.467}          & 0.398          \\ \hline
CodeT5-small$_{ALL}$                 & \multicolumn{1}{c|}{0.824}          & \multicolumn{1}{c|}{0.672}          & \multicolumn{1}{c|}{0.549}          & \multicolumn{1}{c|}{0.457}          & 0.390          \\ \hline
\multirow{2}{*}{\textbf{Model Name}} & \multicolumn{5}{c}{\textbf{Recall@k}}                                                                                                                                  \\ \cline{2-6} 
                                     & \multicolumn{1}{c|}{R@1}            & \multicolumn{1}{c|}{R@2}            & \multicolumn{1}{c|}{R@3}            & \multicolumn{1}{c|}{R@4}            & R@5            \\ \hline
\textbf{CodeT5$_{ALL}$}              & \multicolumn{1}{c|}{\textbf{0.855}} & \multicolumn{1}{c|}{\textbf{0.763}} & \multicolumn{1}{c|}{\textbf{0.732}} & \multicolumn{1}{c|}{\textbf{0.741}} & \textbf{0.765} \\ \hline
DistilBERT$_{ALL}$                   & \multicolumn{1}{c|}{0.835}          & \multicolumn{1}{c|}{0.737}          & \multicolumn{1}{c|}{0.701}          & \multicolumn{1}{c|}{0.707}          & 0.729          \\ \hline
DistilRoBERTa$_{ALL}$                & \multicolumn{1}{c|}{0.830}          & \multicolumn{1}{c|}{0.731}          & \multicolumn{1}{c|}{0.695}          & \multicolumn{1}{c|}{0.701}          & 0.723          \\ \hline
CodeBERT-small$_{ALL}$               & \multicolumn{1}{c|}{0.831}          & \multicolumn{1}{c|}{0.733}          & \multicolumn{1}{c|}{0.699}          & \multicolumn{1}{c|}{0.705}          & 0.727          \\ \hline
CodeT5-small$_{ALL}$                 & \multicolumn{1}{c|}{0.824}          & \multicolumn{1}{c|}{0.723}          & \multicolumn{1}{c|}{0.686}          & \multicolumn{1}{c|}{0.691}          & 0.711          \\ \hline
\multirow{2}{*}{\textbf{Model Name}} & \multicolumn{5}{c}{\textbf{F1-score@k}}                                                                                                                                \\ \cline{2-6} 
                                     & \multicolumn{1}{c|}{F@1}            & \multicolumn{1}{c|}{F@2}            & \multicolumn{1}{c|}{F@3}            & \multicolumn{1}{c|}{F@4}            & F@5            \\ \hline
\textbf{CodeT5$_{ALL}$}              & \multicolumn{1}{c|}{\textbf{0.855}} & \multicolumn{1}{c|}{\textbf{0.726}} & \multicolumn{1}{c|}{\textbf{0.633}} & \multicolumn{1}{c|}{\textbf{0.568}} & \textbf{0.519} \\ \hline
DistilBERT$_{ALL}$                   & \multicolumn{1}{c|}{0.835}          & \multicolumn{1}{c|}{0.702}          & \multicolumn{1}{c|}{0.605}          & \multicolumn{1}{c|}{0.541}          & 0.493          \\ \hline
DistilRoBERTa$_{ALL}$                & \multicolumn{1}{c|}{0.830}          & \multicolumn{1}{c|}{0.696}          & \multicolumn{1}{c|}{0.601}          & \multicolumn{1}{c|}{0.536}          & 0.489          \\ \hline
CodeBERT-small$_{ALL}$               & \multicolumn{1}{c|}{0.831} & \multicolumn{1}{c|}{0.698} & \multicolumn{1}{c|}{0.604} & \multicolumn{1}{c|}{0.540} & 0.492 \\ \hline
CodeT5-small$_{ALL}$                 & \multicolumn{1}{c|}{0.824}   & \multicolumn{1}{c|}{0.689}   & \multicolumn{1}{c|}{0.593}   & \multicolumn{1}{c|}{0.529}   & 0.482   \\ \hline
\end{tabular}
\end{table}


\noindent \textbf{Results and Analysis} 
The performance of smaller PTMs under the \toolname framework is summarized in Table \ref{tab:rq4}. 
DistilBERT gives the best performance. 
The worst-performing variant of smaller PTMs is CodeT5-small$_{ALL}$, where it achieves scores of 0.824, 0.689, 0.593, 0.529, and 0.482 with respect to $F1$-$score@1$-$5$.
We also observe that these smaller variants all can outperform the previous state-of-the-art method, Post2vVec, by a substantial margin. 

Inference latency refers to the time taken for each model to make predictions. Generally speaking, inference latency is affected by the computing power of the running machine and the length of the input sequence. To make a fair comparison, we adopt the same hardware to query the models. To be specific, we use two Nvidia Tesla v100 16GB GPUs to run the model. 
As specified in Section \ref{sec:experiment}, we set the input lengths to be the same as in training, which is 100 for Title, 512 for Description, and 512 for Code. We randomly sample 2,000 examples from the test set and calculate the average inference latency taken by the models. To further reduce the effects of randomness, the experiments are repeated five times. 
s

\begin{figure}[ht]
    
	\includegraphics[width=1\linewidth]{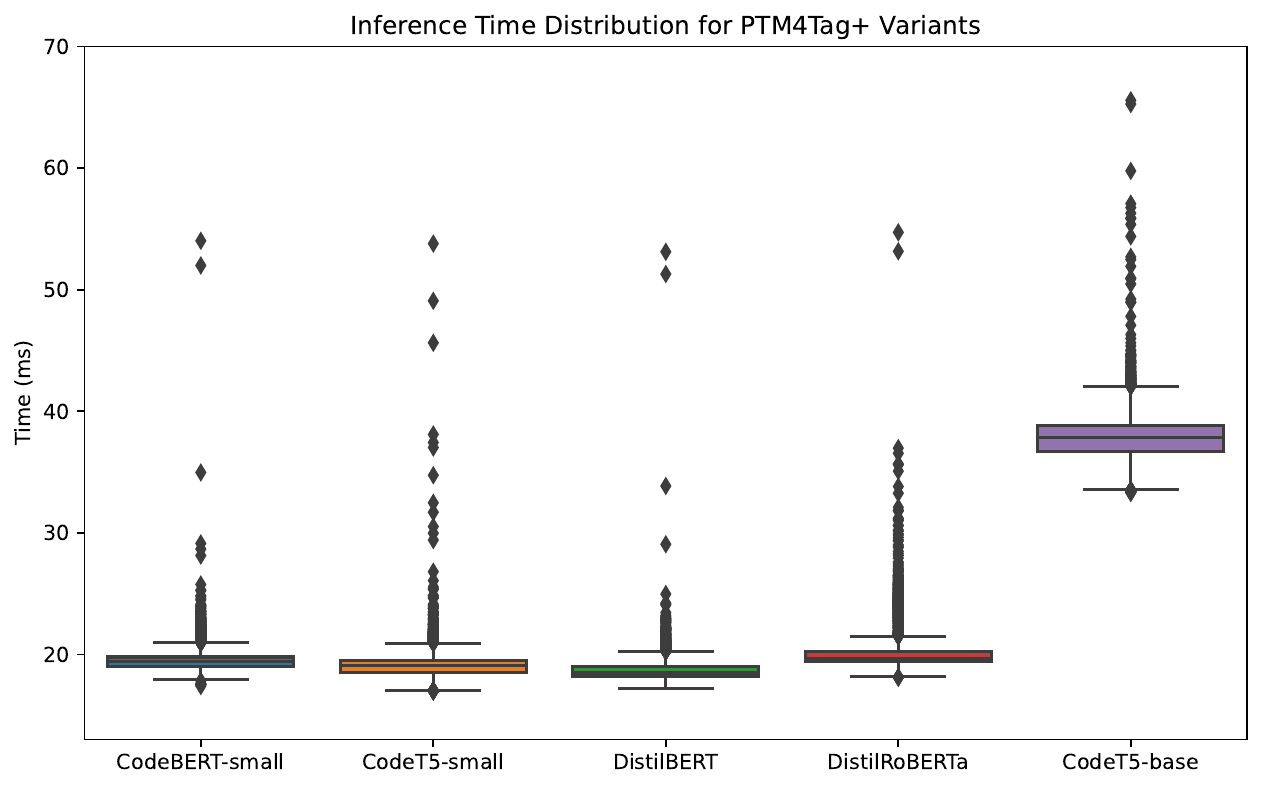}
 
	\caption{A box-plot demonstrate the distribution of inference time of CodeT5$_{ALL}$ and small variants of \toolname among 10,000 samples.}
 \label{fig:time}
\end{figure}

Table \ref{tab:small} summarizes the inference time improvement and performance drop of smaller PTMs compared with CodeT5 under the \toolname framework. On average, the inference latency is reduced by over 47.2\% while at least 93.96\% of the original performance could be preserved in terms of average $F1$-$score@k$. 
In Table \ref{tab:time-stats}, we present a detailed statistical summary of the inference time (measured in milliseconds) for CodeT5$_{ALL}$ and several smaller variants of \toolname, based on a sample size of 10,000 (2,000 $\times$ 5). Figure \ref{fig:time} presents the boxplot on the distribution of inference time. The CodeT5-base model exhibited the highest average inference time at 37.8 ms. DistilBERT and DistilRoBERTa demonstrated more moderate average inference times of 18.6 ms and 20.0 ms, respectively. The CodeBERT-small and CodeT5-small models yielded similar results, with mean inference times of 19.5 ms and 19.1 ms, respectively. From Table \ref{tab:time-stats}, we can clearly see that CodeT5$_{ALL}$ has the highest inference time, with a median of 37.9ms and a maximum of 114.7ms. In contrast, the smaller models, including DistilBERT, DistilRoBERTa, CodeBERT-small, and CodeT5-small, have notably faster inference times, with medians ranging from 18.5ms to 19.7ms. Among these, CodeT5-small has the shortest minimum inference time at 16.9ms. The standard deviation values suggest that the inference times for these models are relatively consistent, with CodeT5-base having the most variability.

Our results demonstrate that smaller variants of \toolname could outperform larger models like BERTOverflow$_{ALL}$, PLBART$_{ALL}$. Such a phenomenon suggests the performance of a model does not increase linearly with its size in the SO post tagging task and it implies parameter redundancy of large models. Much previous literature showed that smaller PTMs could also give competent performance and the performance gap to larger PTMs is insignificant~\citep{adoma2020comparative,giorgi2020declutr} and sometimes they are even better when the training is carefully conducted~\citep{sarfraz2021knowledge}. For example, Wang et al. claimed that CodeT5-small yields better performance than PLBART with a smaller size in tasks like code summarization and code generation~\citep{codet5}. For a practical tool, the trade-off between inference latency and performance should be cautiously determined. 

\begin{table}[ht]
\centering
\caption{Statistical summary for distribution of inference time (ms) of CodeT5$_{ALL}$ and small variants of \toolname among 10,000 samples.}
\label{tab:time-stats}
\begin{tabular}{c|ccccccc}
\hline
\textbf{Model Name}  & \textbf{std} & \textbf{min} & \textbf{25\%} & \textbf{50\%} & \textbf{75\%} & \textbf{max} \\ \hline
\textbf{CodeT5-base}             & 2.0         & 33.3        & 36.7         & 37.9         & 38.9         & 114.7       \\
\textbf{DistilBERT}              & 0.8         & 17.2        & 18.2         & 18.5         & 19.0         & 53.1        \\
\textbf{DistilRoBERTa}           & 1.4         & 18.1        & 19.4         & 19.7         & 20.3         & 54.7        \\
\textbf{CodeBERT-small}          & 0.9         & 17.4        & 19.0         & 19.4         & 19.8         & 54.0        \\
\textbf{CodeT5-small}         & 1.1         & 16.9        & 18.5         & 19.1         & 19.5         & 53.8        \\
\hline
\end{tabular}
\end{table}

\begin{table}[]
\centering
\caption{Comparison of smaller PTMs under \toolname framework with the best-performing model, CodeT5$_{ALL}$, including the mean inference time(ms), inference time improvement(\%), and F1-Score Performance drop(\%). }
\label{tab:small}
\begin{tabular}{c|c|ccccc}
\hline
\multirow{2}{*}{\textbf{Model Name}} & \multirow{2}{*}{\textbf{\begin{tabular}[c]{@{}c@{}}Inference \\ Latency(ms)\end{tabular}}} & \multicolumn{5}{c}{\textbf{F1-Score Performance Drop}}                                                                                                            \\ \cline{3-7} 
                                     &                                                                                            & \multicolumn{1}{c|}{\textbf{F1@1}} & \multicolumn{1}{c|}{\textbf{F1@2}} & \multicolumn{1}{c|}{\textbf{F1@3}} & \multicolumn{1}{c|}{\textbf{F1@4}} & \textbf{F1@5} \\ \hline
DistilBERT$_{ALL}$                   & 18.6(-50.9\%)                                                                              & \multicolumn{1}{c|}{-2.12\%}       & \multicolumn{1}{c|}{-3.20\%}       & \multicolumn{1}{c|}{-3.84\%}       & \multicolumn{1}{c|}{-4.46\%}       & -4.68\%       \\ \hline
DistilRoBERTa$_{ALL}$                & 20.0(-47.2\%)                                                                              & \multicolumn{1}{c|}{-1.53\%}       & \multicolumn{1}{c|}{-2.36\%}       & \multicolumn{1}{c|}{-3.2\%}        & \multicolumn{1}{c|}{-3.57\%}       & -3.90\%       \\ \hline
CodeBERT-small$_{ALL}$               & 19.5(-48.5\%)                                                                             & \multicolumn{1}{c|}{-2.00\%}       & \multicolumn{1}{c|}{-2.92\%}       & \multicolumn{1}{c|}{-3.36\%}       & \multicolumn{1}{c|}{-3.74\%}       & -4.09\%       \\ \hline
CodeT5-small$_{ALL}$                 & 19.1(-49.6\%)                                                                              & \multicolumn{1}{c|}{-2.83\%}       & \multicolumn{1}{c|}{-4.17\%}       & \multicolumn{1}{c|}{-5.12\%}       & \multicolumn{1}{c|}{-5.70\%}       & -6.04\%       \\ \hline
CodeT5$_{ALL}$                       & \textbf{37.8}                                                                              & \multicolumn{1}{c|}{-}             & \multicolumn{1}{c|}{-}             & \multicolumn{1}{c|}{-}             & \multicolumn{1}{c|}{-}             & -             \\ \hline
\end{tabular}
\end{table}

\begin{tcolorbox}
    \textbf{Answers to RQ4}: 
    Using smaller PTMs under \toolname, the inference latency is reduced by over 47.2\% on average while at least 93.96\% of the original performance could be preserved in terms of average $F1$-$score@k$. 
\end{tcolorbox}

\section{Discussion}
\label{sec:discussion}

\subsection{\textbf{Error Analysis}}
We conduct an error analysis to illustrate the capability of our proposed framework \toolnamenospace. Take a Stack Overflow post\footnote{\url{https://stackoverflow.com/questions/51910978}} titled \emph{Pass Input Function to Multiple Class Decorators} in the test dataset as an example.
The ground truth tags of the post are {\tt decorator}, {\tt memoization}, {\tt python}, {\tt python-2.7}, and {\tt python-decorators}. The tags predicted by Post2Vec are {\tt class}, {\tt timer}, {\tt python-decorators}, {\tt decorator}, and {\tt  python}
 while \toolname gives the exact prediction as to the ground truth tags. Although the word \emph{memoization} has occurred several times in the post, Post2Vec still failed to capture its existence and the connection between \emph{memoization} and \emph{decorator}. Moreover, we found that the CodeSearchNet database \citep{codesearchnet} which is used to pre-train CodeT5 includes source code files that relate to both \emph{memoization} and \emph{decorator}.\footnote{\url{https://github.com/nerdvegas/rez/blob/1d3b846d53b5b5404edfe8ddb9083f9ceec8c5e7/src/rez/utils/memcached.py\#L248-L375}} This potentially could indicate that the pre-trained knowledge learned by CodeT5 is beneficial for our task.

From the 100,000 posts in the test data, 2,707 posts yield an $F1$-$score@5$ of zero. To comprehend which tags are less effectively recognized by \toolnamenospace, we count the occurrences of ground truth tags associated with posts that received an $F1$-$score@5$ of zero. Table \ref{tab:error-analysis} presents the top-10 most often missed
tags of this analysis, showing that \textit{python} and \textit{python-3.x} are the tags most frequently missed by \toolnamenospace. We notice that this is because \toolname cannot accurately handle the version of python; it may likely produce the prediction of \textit{python-2.7}.

\begin{table}[]
\centering
\caption{Top-10 Tags Most Often Missed in Posts with an $F1$-$score@5$ of 0.}
\label{tab:error-analysis}
\begin{tabular}{c|c}
\hline
           & Frequency \\ \hline
python-3.x & 58        \\ \hline
python     & 45        \\ \hline
javascript & 37        \\ \hline
c\#         & 29        \\ \hline
java       & 29        \\ \hline
php        & 22        \\ \hline
angular    & 22        \\ \hline
android    & 20        \\ \hline
flutter    & 18        \\ \hline
web        & 18        \\ \hline
.net       & 17        \\ \hline
\end{tabular}%

\end{table}

\subsection{\textbf{Manual Evaluation}}
Furthermore, we conducted a manual evaluation of the predicted tags of \toolnamenospace. We randomly sample a statistically representative subset of 166 posts in our testing dataset which gives us a confidence level of 99\% with a confidence interval of 10\footnote{we use the sample size calculator from \url{https://www.surveysystem.com/sscalc.htm}}. We invite three software developers with at least 5 years of programming experience to evaluate the relevancy of the tags predicted by \toolnamenospace. For each post, we present the top-5 tags predicted by \toolnamenospace. Each of the developers is required to annotate 166 posts individually, thus 166 x 5 = 830 tags in total. Each annotator is required to use domain expertise and is allowed to refer to external sources when checking the relevance of the predicted tags. Out of the 166 posts, the tags predicted for 76 posts are marked as containing no irrelevant tags by all three annotators. We then asked the three annotators to discuss their evaluation results. After the discussion, the three annotators agreed that there were a total of 49 posts containing 62 irrelevant tags. 

From our manual analysis, we observed that a significant majority of the predicted tags are related to the post. However, some inaccuracies were also noted. We found that most of the irrelevant predictions of \toolname are not completely wrong. Instead, these irrelevant predictions are indirectly related to the topic of the post. We present six examples in Table 11. For example, for post 52109809, the predicted tags are: \texttt{bitbucket}, \texttt{git}, \texttt{git-config}, \texttt{github}, \texttt{ssh}. However, the post is solely about \texttt{git} and is not related to \texttt{bitbucket}, although \texttt{bitbucket} and \texttt{github} are related topics. Another noticeable observation is the simultaneous predictions of tags such \texttt{python}, \texttt{python-3.x}, and \texttt{python 2.7} always come together, which also outlines a limitation of our approach. 

Further, we found the \toolname yielded wrong predictions in certain niche topics. For example, for post 51903596, \toolname gives the prediction of \texttt{ruby} and \texttt{julia-lang}, while the question is about the crystal programming language.

\begin{table}[]
\centering
\caption{Examples of Predicted Tags of PTM4Tag+}
\label{tab:manual-analysis}
\resizebox{\textwidth}{!}{%
\begin{tabular}{c|c|c}
\hline
\textbf{Post ID} & \textbf{Title}                                                                                                                                           & \textbf{Predicted Tags}                                                                                          \\ \hline
51903596         & how to resolve generics in macros?                                                                                                                       & \begin{tabular}[c]{@{}c@{}}elixir, julia-lang, \\ macros, metaprogramming, \\ ruby\end{tabular}      \\ \hline
52109809         & Same user set after changing username globally to git                                                                                                    & \begin{tabular}[c]{@{}c@{}}bitbucket, git, \\ git-config, github, \\ ssh\end{tabular}                \\ \hline
52018972         & \begin{tabular}[c]{@{}c@{}}using else in basic program \\ getting error: unindent does not \\ match any outer indentation level\end{tabular}             & \begin{tabular}[c]{@{}c@{}}if-statement, indentation, \\ python, python-2.7, python-3.x\end{tabular} \\ \hline
51929049         & \begin{tabular}[c]{@{}c@{}}after running the above code why \\ i am getting just one output for \\ type () function? why not for all three?\end{tabular} & \begin{tabular}[c]{@{}c@{}}function, python, \\ python-2.7, python-3.x, types\end{tabular}           \\ \hline
51905926         & \begin{tabular}[c]{@{}c@{}}python how to create a string \\ that is 4 characters long from a number\end{tabular}                                         & \begin{tabular}[c]{@{}c@{}}integer, python, \\ python-2.7, python-3.x, string\end{tabular}           \\ \hline
\end{tabular}%
}
\end{table}

\subsection{\textbf{Revised Experiments for Updated Dataset}}

To further assess the efficacy of \toolnamenospace, we conduct the evaluation of \toolname on an augmented dataset sourced from the most recent Stack Overflow dump dated June 2023.\footnote{\url{https://archive.org/details/stackexchange}}. Following the preprocessing procedures we mentioned in Section \ref{sec:experiment}, tags with occurrences fewer than 50 times were filtered out. This updated dataset comprises 23,687 common tags and 22,498,254 posts. We use 22,398,254 posts as the training set and the latest 100,000 posts as the test set. We train CodeT5$_{ALL}$ on the updated dataset, which is the variant that has the highest mean F1-score@5 on the previous dataset. The experiment is conducted in the same setting we mentioned in Section \ref{sec:experiment}. The experimental result is demonstrated in Table \ref{tab:codet5-new}. Overall, CodeT5$_{ALL}$ yields an $F1$-$score@5$ of 0.516 on the latest data, which is comparable to 0.519 from the previous dataset. 

\begin{table}[h]
\centering
\caption{Experiment Results for CodeT5$_{ALL}$ on the Updated Dataset}
\label{tab:codet5-new}
\begin{tabular}{c|ccccc}
\hline
\multirow{2}{*}{\textbf{Model Name}} & \multicolumn{5}{c}{\textbf{Precision@k}}                                                                                                                               \\ \cline{2-6} 
                                     & \multicolumn{1}{c|}{P@1} & \multicolumn{1}{c|}{P@2} & \multicolumn{1}{c|}{P@3} & \multicolumn{1}{c|}{P@4} & P@5 \\ \hline
\textbf{CodeT5$_{ALL}$}              & \multicolumn{1}{c|}{0.839} & \multicolumn{1}{c|}{0.703} & \multicolumn{1}{c|}{0.585} & \multicolumn{1}{c|}{0.494} & 0.424 \\ \hline

\multirow{2}{*}{\textbf{Model Name}} & \multicolumn{5}{c}{\textbf{Recall@k}}                                                                                                                                  \\ \cline{2-6} 
                                     & \multicolumn{1}{c|}{R@1} & \multicolumn{1}{c|}{R@2} & \multicolumn{1}{c|}{R@3} & \multicolumn{1}{c|}{R@4} & R@5 \\ \hline
\textbf{CodeT5$_{ALL}$}              & \multicolumn{1}{c|}{0.839} & \multicolumn{1}{c|}{0.752} & \multicolumn{1}{c|}{0.716} & \multicolumn{1}{c|}{0.722} & 0.744 \\ \hline

\multirow{2}{*}{\textbf{Model Name}} & \multicolumn{5}{c}{\textbf{F1-score@k}}                                                                                                                               \\ \cline{2-6} 
                                     & \multicolumn{1}{c|}{F@1} & \multicolumn{1}{c|}{F@2} & \multicolumn{1}{c|}{F@3} & \multicolumn{1}{c|}{F@4} & F@5 \\ \hline
\textbf{CodeT5$_{ALL}$}              & \multicolumn{1}{c|}{0.839} & \multicolumn{1}{c|}{0.719} & \multicolumn{1}{c|}{0.627} & \multicolumn{1}{c|}{0.563} & 0.516 \\ \hline
\end{tabular}
\end{table}

\subsection{\textbf{Threats to Validity}}

\subsubsection*{\textbf{Threats to internal validity}} To ensure we implement the baseline (i.e., Post2Vec) correctly, we reused the official replication package released by the Xu et al.\footnote{\url{https://github.com/maxxbw54/Post2Vec}} To instantiate variants of \toolname with different pre-trained models, we utilized a widely-used deep learning library \textit{Hugging Face}.\footnote{\url{https://huggingface.co/}} Similar to prior studies~\citep{tagcnn,tagcombine,post2vec}, our work assumes that the tags are labeled correctly by users in Stack Overflow. However, some tags are potentially mislabelled. Still, we believe that Stack Overflow’s effective crowdsourcing process helps to reduce the number of such cases, and we further minimize this threat by discarding rare tags and posts (as described in Section~\ref{subsec:data_preparatin}). We conducted a manual analysis to assess the relevancy of the tags of Stack Overflow posts. We computed a statistically representative sample size using a popular sample size calculator\footnote{\url{https://www.surveysystem.com/sscalc.htm}} with a confidence level of 99\% and a confidence interval of 10. We sampled 166 code snippets to conduct the manual evaluation where three experienced developers with at least 5 years of programming background were involved in this manual evaluation. They individually assessed the relevance of the tags, leveraging both their expertise and external sources. A discussion was then held among the three evaluators if they identified any conflicts. They all agreed that the tags associated with the sampled posts were relevant to the content. Another threat to the internal validity is the hyperparameter setting we used to fine-tune \toolnamenospace. To mitigate this threat, we use hyper-parameters that were reported in prior studies as recommended or optimal~\citep{bert, CodeBERT, wang2022training}.

Users of SO can put any kind of text into the \textit{Code} blocks of SO posts. The \textit{Code} component of \toolname may contain other types of content than code snippets, such as stack traces and error messages. We refer to these content as non-code content. To study the impact of non-code contents on our framework, we randomly sample a statistically representative subset of 166 posts from our test dataset, providing us with a 99\% confidence level and a 10\% confidence interval. On manual inspection of this subset, we discern that 36 posts, which is 21.7\% of the subset, have content other than typical code snippets within the \textit{Code} component. Breaking it down, 22 posts featured error messages, 5 posts detailed database schema layouts, 5 posts delineated input/output format descriptions, and 4 posts showcased actual program outputs. The performance of \toolname (CodeT5$_{ALL}$) in this sampled subset and the posts containing non-code content are presented in Table \ref{tab:noncode}. Notably, the average \(F1\)-\(score@5\) for posts with non-code content surpasses that of the overall subset. This suggests that \toolname has the potential to not only deal with typical code snippets but also with other forms of content within the \textit{Code} component.

\begin{table}[]
\centering
\caption{Comparison of \toolname average performance on a statistically representative subset from the test set versus \toolname average performance on Non-code Text within the same subset. P represents Precision, R represents Recall, and F represents the F1-Score. }
\label{tab:noncode}
\begin{tabular}{c|c|c|c|c|c}
\hline
 & \textbf{P@1} & \textbf{P@2} & \textbf{P@3} & \textbf{P@4} & \textbf{P@5} \\ \hline
\textbf{Subset } & 0.867 & 0.732 & 0.604 & 0.503 & 0.435 \\ \hline
\textbf{Non-code} & 0.886 & 0.771 & 0.629 & 0.521 & 0.446 \\ \hline
 & \textbf{R@1} & \textbf{R@2} & \textbf{R@3} & \textbf{R@4} & \textbf{R@5} \\ \hline
\textbf{Subset} & 0.867 & 0.792 & 0.742 & 0.735 & 0.769 \\ \hline
\textbf{Non-code} & 0.886 & 0.786 & 0.705 & 0.698 & 0.720 \\ \hline
 & \textbf{F@1} & \textbf{F@2} & \textbf{F@3} & \textbf{F@4} & \textbf{F@5} \\ \hline
\textbf{Subset} & 0.867 & 0.752 & 0.647 & 0.573 & 0.530 \\ \hline
\textbf{Non-code} & 0.886 & 0.776 & 0.656 & 0.583 & 0.535 \\ \hline
\end{tabular}
\end{table}

\subsubsection*{\textbf{Threats to external validity}} 
We analyzed Stack Overflow, the largest SQA site, with a massive amount of questions. These questions cover diverse discussions on various software topics. As software technologies evolve fast, our results may not generalize to those newly emerging topics. Instead, our framework can adapt to new posts by fine-tuning models on more and new questions.

\subsubsection*{\textbf{Threats to construct validity}} 
Threats to construct validity are related to the suitability of our evaluation metrics. $Precision@k$, $Recall@k$, and $F1$-$score@k$ are widely used to evaluate many tag recommendation approaches in software engineering~\citep{tagcnn,tagcombine,post2vec}. Thus, we believe the threat is minimal.
We reuse the evaluation metrics proposed in our baseline method, Post2Vec~\citep{post2vec}. 
We conduct the Wilcoxon signed-rank statistical hypothesis test~\citep{gehan1965generalized}
on the paired data which corresponds to the F1-score@5 of CodeT5$_{ALL}$ and all other PTM models. We conducted the Wilcoxon Signed Rank Test at a 95\% confidence level (i.e., p-value $<$ 0.05). 
We found that CodeT5$_{ALL}$ significantly outperforms all other variants with a threshold of \( p < 0.05 \). In addition, we conducted Cliff's delta~\citep{cliff1993dominance} to measure the effect size of our results. The Cliff's Delta statistic,  denoted as $|\delta|$, is a non-parametric effect size measure that quantifies the amount of difference between two groups of observations beyond p-values interpretation. 
We consider $|\delta|$ that are classified as ``Negligible (N)" for $|\delta| < 0.147$, ``Small (S)" for $0.147 \leq |\delta| < 0.33$, ``Medium (M)" for $0.33 \leq |\delta| < 0.474$, and ``Large (L)" for $|\delta| \geq 0.474$,  respectively following previous literature~\citep{cliff2014ordinal}. We observe that CodeT5$_{ALL}$ substantially outperforms the baseline models – Cliff’s deltas are not negligible and are in the range of 0.21 (small) to 0.54 (large) .


\subsection{\textbf{Lessons Learned}}
\label{subsec:lessons}

\subsubsection*{\textbf{Lesson \#1}
\textbf{\textit{Pre-trained language models are effective in tagging SO posts.}}}
The tag recommendation task for SQA sites has been extensively studied in the last decade~\citep{tagdc, tagcnn, tagcombine, post2vec}. Researchers have tackled the problem via a range of techniques, e.g., collaborative filtering~\citep{tagdc} and deep learning~\citep{post2vec}. Furthermore, these techniques usually involve separate training for each component. Our experiment results have demonstrated that simply fine-tuning the pre-trained Transformer-based model can achieve state-of-the-art performance, even if there are thousands of tags. CodeBERT, BERT, and RoBERTa are capable of providing promising results for tag recommendation. Even though BERT and RoBERTa did not leverage programming language at the pre-training stage. 

We encourage practitioners to leverage pre-trained models in the {\em multi-label} classification settings where the size of the label set could go beyond thousands. Although PTMs are already widely adopted in SE tasks, most tasks are formulated as either {\em binary} classification problems or {\em multi-class} classification problems. In binary or multi-class classification problems the label classes are mutually exclusive, whereas for multi-label classification problems, each data point may belong to several labels simultaneously. Moreover, our experiments also validate the generalizability of pre-trained models. We recommend practitioners apply pre-trained models in more SE tasks and consider fine-tuning pre-trained models as one of their baselines. 

\subsubsection*{\textbf{Lesson \#2}}
\textbf{\textit{Encoder-decoder models should also be considered in SO-related tasks for classification tasks.}} 
By convention, BERT-based models are widely used for classification tasks in generating sentence embeddings. 
Our results have shown that encoder-decoder models are also capable of generating meaningful embeddings (especially CodeT5 gives the best performance) in the tag recommendation task of SO posts. We advocate researchers also involve the encoder-decoder models as baseline methods for SO-related classification tasks in the future.

\subsubsection*{\textbf{Lesson \#3}}
\textbf{\textit{All components of a post from Stack Overflow are valuable pieces of semantics.}} Most previous literature has removed the code snippets from the pre-training process because they are considered noisy, poorly structured, and written in many different programming languages~\citep{tagcnn, tagdc, tagcombine, tagmulrec}. However, our results show that code snippets are also beneficial to capture the semantics of SO posts and further boost the performance of the tag recommendation task. We encourage researchers to consider both the natural and programming language parts of a post when analyzing SQA sites.

\subsubsection*{\textbf{Lesson \#4}}
\textbf{\textit{Smaller pre-trained models are practical substitutes.}}
We demonstrate that various small PTMs could achieve similar performance to larger PTMs while increasing the inference latency.
Smaller variants of \toolname even outperformed variants with BERTOverflow and PLBART. We show that smaller PTMs are also effective in the considered task, and developers should consider these PTMs to reach a balance point between the performance and usability in the real world for SO-related tasks.

\section{Related Work}

\label{sec:related_work}
\subsection{Pre-trained Models}
Transformer-based pre-trained models have recently benefited a broad range of both understanding and generation tasks.
Recent works~\citep{biobert,scibert,clinicalbert} have shown that the in-domain knowledge acquired by PTMs is valuable in improving the performance on domain-specific tasks, such as ClinicalBERT~\citep{clinicalbert} for clinical text, SciBERT~\citep{scibert} for scientific text, and BioBERT \citep{biobert} for biomedical text. Evoked by the success of PTMs in other domains, researchers have started to work on the SE domain-specific PTMs~\citep{CodeBERT, sebert, cbert, cubert, bertoverflow}. Developers are free to create and pick arbitrary identifiers. These identifiers introduce a lot of customized words into the texts within the SE field~\citep{shi2022identifier}. This suggests that models pre-trained on general texts, such as BERT~\citep{bert}, may not be optimal for representing texts in the software engineering domain.

Specifically, the transformer model is designed with an encoder-decoder architecture. The encoder takes an input sentence and derives important features from it, while the decoder leverages these features to generate an output sentence. In terms of architecture, we categorize the transformer-based pre-trained models into three types, which are encoder-only models, decoder-only models, and encoder-decoder models

Encoder-only pre-traiend models  are essentialy the encoder part of the transformer architectures. Encoder-only models are widely generating sentence representations in language understanding tasks\citep{lan2020self}. 
Buratti et al. trained C-BERT~\citep{cbert} using code from the top-100 starred GitHub C language repositories. Experimental results show that C-BERT achieves high accuracy in the Abstract Syntax Tree (AST) tagging task and produces comparatively good performance to graph-based approaches on the software vulnerability identification task.
Von der Mosel et al.~\citep{sebert} aims to provide a better SE domain-specific pre-trained model than BERTOverflow~\citep{bertoverflow} and propose seBERT~\citep{sebert}. Since seBERT is developed upon the BERT$_{LARGE}$ architecture, we did not investigate the effectiveness of seBERT under the \toolname framework due to restraints on computational resources and GPU memory consumption. 
Guo et al. presented GraphCodeBERT~\citep{graphcodebert}, the first pre-trained models which consider the inherent structure of programming languages. In addition to NL and PL data, GraphCodeBERT also involves data flow information. In the pre-training stage, Guo et al. utilize masked language modeling and two new structure-aware tasks as pre-training objectives. We did not include GraphCodeBERT in our work because it is difficult to obtain data flow graphs for the code snippets within a SO post.

Decoder-only pre-trained models only inherit the decoder part of the transformer architecture. Decoder-only models are usually used for generation tasks \citep{Wang2022TowardsUC}. 
Svyatkovskiy et al. trained GPT-C~\citep{gpt-c}, a generative model based on GPT-2 architecture, and leveraged GPT-C to build a code completion tool.

Encoder-decoder pre-trained models use complete Transformer architecture.
PLBART~\citep{plbart} has undergone pre-training on a massive set of Java and Python functions and corresponding natural language text through the denoising autoencoding process. Results showed that PLBART is promising in code summarization, code generation, and code translation tasks. 
Wang et al. introduced CodeT5~\citep{codet5}, a pre-trained encoder-decoder Transformer model that effectively utilizes the semantic information conveyed through developer-assigned identifiers. Extensive experiments demonstrate that CodeT5 significantly outperforms previous methods in tasks such as code defect detection and clone detection, as well as generation tasks including PL-NL, NL-PL, and PL-PL translations.
CoTexT also adopted the T5 architecture~\citep{cotext}.  CoTexT is another bi-model PTM and is capable of supporting a range of natural language-to-programming language tasks, including code summarization and documentation, code generation, defect detection, and debugging. 
TreeBERT is a tree-based pre-trained model for improving code-related generation tasks, proposed by Jiang et al.~\citep{treebert}.  TreeBERT leverages the abstract syntax tree (shortened as AST) corresponding to the code into consideration. The model is pre-trained by the tree masked language modeling (TMLM) task and node order prediction (NOP) task. As the result, TreeBERT achieved state-of-the-art performance in code summarization and code documentation tasks.

\subsection{Tag Recommendation for SQA Sites}
Researchers have already extensively studied the tag recommendation task in the SE domain and proposed a number of approaches.
Wang et al.~\citep{tagcombine} proposed TagCombine, a tag recommendation framework that consists of three components: a multi-label ranking component, a similarity-based ranking component, and a tag-term-based ranking component. Eventually, TagCombine leverages a sample-based method to combine the scores from the three components linearly as the final score.
Wang et al. introduced EnTagRec~\citep{entagrec} that utilizes Bayesian inference and an enhanced frequentist inference technique. Results show that it outperformed TagCombine by a significant margin. Wang et al. then further extends EnTagRec~\citep{entagrec} to EnTagRec++~\citep{entagresplusplus}, the latter of which additionally considers user information and an initial set of tags provided by a user. Zhou et al. proposed TagMulRec~\citep{tagmulrec}, a collaborative filtering method that suggests new tags for a post based on the results of semantically similar posts. 
Li et al. proposed TagDC~\citep{tagdc}, which is implemented with two parts: TagDC-DL that leverages a content-based approach to learn a multi-label classifier with a CNN Capsule network, and TagDC-CF which utilizes collaborative filtering to focus on the tags of similar historical posts. 
Post2Vec~\citep{post2vec} distributively represents SO posts and is shown to be useful in tackling numerous Stack Overflow-related downstream tasks. We select Post2Vec~\citep{post2vec} as the baseline in our experiments as it achieves the state-of-the-art performance in the tag recommendation task for SO posts.
The current state-of-the-art approach, Post2Vec \citep{post2vec} achieves the state-of-the-art performance in the tag recommendation task of SO, and it is also capable of generating the distributed representation of SO posts that could be useful in numerous Stack Overflow-related downstream tasks. 
\section{Conclusion and Future Work}
\label{sec:conclusion}

In this work, we introduce \toolnamenospace, a pre-trained model-based framework for tag recommendation of Stack Overflow posts.
We implement eight variants of \toolname with different PTMs. Our experiment results show that the CodeT5 gives the best performance under the framework of \toolnamenospace, and it outperforms the state-of-the-art approach by a large margin in terms of $F1$-$score@5$. 
However, \toolname variants implemented with BERTOverflow and ALBERT do not give promising results. Even though PTMs are shown to be powerful and effective, PTMs behave differently, and the selection of PTMs needs to be carefully decided.

In the future, we are interested in applying \toolname on more SQA sites such as AskUbuntu\footnote{\url{https://askubuntu.com/}}, etc., to evaluate its effectiveness and generalizability further. It is noteworthy that beyond actual code snippets, SO posts often incorporate other textual artifacts like stack traces and error messages. We plan to expand our approach to more fine-grained types of text for future research.

\section*{Data Availability}

We release our replication package to facilitate future research. The code and datasets used during the study are available in the \url{10.6084/m9.figshare.24268831}.

\section*{Declarations}
The authors have no conflicts of interest/competing Interests to declare that are relevant to this paper.

\bibliography{reference}
\end{document}